\documentclass[review]{elsarticle}
\usepackage{lineno,natbib}
\journal{International Journal of Multiphase Flow}

\usepackage{graphicx}
\usepackage{amsfonts}
\usepackage{amssymb}
\usepackage{float}
\usepackage{latexsym}
\usepackage{amsmath}
\usepackage{placeins}
\usepackage{caption}
\usepackage{subcaption}
\usepackage[euler]{textgreek}
\usepackage{textcomp}
\usepackage{siunitx}
\usepackage{graphics}
\usepackage{multirow}
\usepackage{tabularx}
\usepackage[title]{appendix}
\usepackage{booktabs}

\usepackage[table]{xcolor}

\newcommand{\tcr}[1]{\textcolor{black}{#1}}

\newcommand{\eqr}[1]{(\ref{eq:#1})}

\newcommand{\ods}[2]{\frac{d^2 #1}{d #2^2}}

\newcommand{\pds}[2]{\frac{\partial^2 #1}{\partial #2 ^2}}
\newcommand{\eg}{\textit{e.g.}}

\newcommand{\etal}{\textit{et al.}}
\newcommand{\ie}{\textit{i.e.}}

\begin{document}
	\begin{frontmatter}
		
		\title{Effect of gas viscosity on the interfacial instability development in a two-phase mixing layer}
		
		\author[add1]{Tanjina Azad}
		\author[add1]{Yue Ling\corref{cor1}}
		\ead{Stanley\_Ling@sc.edu}
		\cortext[cor1]{Corresponding author. } 
		\address[add1]{Department of Mechanical Engineering, University of South Carolina, 300 Main St, Columbia, SC 29208, USA}


\begin{abstract}
The interfacial instability in a two-phase mixing layers between parallel gas and liquid streams is important to two-phase atomization. Depending on the inflow conditions and fluid properties, interfacial instability can be convective or absolute. The goal of the present study is to investigate the impact of gas viscosity on the interfacial instability. Both interface-resolved simulations and linear stability analysis (LSA) have been conducted. In LSA, the Orr-Sommerfeld equation is solved to analyze the spatio-temporal viscous modes. When the gas viscosity decreases, the Reynold number ($\text{Re}$) increases accordingly. The LSA demonstrates that when $\text{Re}$ is higher than a critical threshold, the instability transitions from the absolute to the convective (A/C) regimes. Such a $\text{Re}$-induced A/C transition is also observed in the numerical simulations, though the critical Re observed in simulations is significantly lower than that predicted by LSA. The LSA results indicate that the temporal growth rate decreases with Re. When the growth rate reaches zero, the A/C transition will occur. The $\text{Re}$-induced A/C transition is observed in both confined and unconfined mixing layers and also in cases with low and high gas-to-liquid density ratios. In the transition from typical absolute and convective regimes, a weak absolute regime is identified in the simulations, for which \tcr{the spectrograms} show both the absolute and convective modes. The dominant frequency in the weak absolute regime can be influenced by the perturbation introduced at the inlet. The simulation results also show that the wave propagation speed can vary in space. In the absolute instability regime, the wave propagation speed agrees well with the absolute mode celerity near the inlet and increases to the Dimotakis speed further downstream. 
\end{abstract}

\begin{keyword}
	Two-phase mixing layer \sep  Interfacial instability \sep Viscous effect
\end{keyword}

\end{frontmatter}


\section{Introduction}
\label{sec:intro}
Two-phase mixing layers play an essential role in the spray formation through air-blast or air-assisted atomization \cite{Lefebvre_1980a, Lasheras_1998a, Varga_2003a, Lefebvre_2017a}. When the parallel gas and liquid steams meet at the end of the separator plate, the velocity difference triggers a shear instability on the interface. The interfacial instability can be convective or absolute \cite{Otto_2013a}, depending on the inflow conditions and the fluid properties. The dynamic pressure ratio has been shown to be an important parameter to determine the convective to absolute (C/A) transition \cite{Fuster_2013a}. Previous linear stability analysis showed that the surface tension plays a significant role on the instability and the selection of the most unstable mode \cite{Otto_2013a}. The presence of the separator plate between the two streams will induce a velocity deficit in the inlet velocity profile, which is also found to promote the C/A transition. Both inviscid and viscous stability analysis have been conducted, while the inviscid theory well predict the scaling relation of the most unstable wavelength and frequencies \cite{Rangel_1988a, Raynal_1997a, Marmottant_2004a}, the viscous stability analysis is needed to yield accurate prediction of the magnitudes of the most unstable frequency and wavelength \cite{Fuster_2013a, Otto_2013a, Matas_2015a}. Nevertheless, there remains discrepancy in the most-unstable frequencies predicted by viscous instability analysis and experiments \cite{Otto_2013a}. 

Spatial-temporal linear stability analysis (LSA) has been used to investigate the C/A transition \cite{Otto_2013a, Matas_2015a, Matas_2018a}. The instability becomes absolute when the spacial branches on the wavenumber plane pinch, forming a saddle point \cite{Briggs_1964a}. Two distinct absolute instabilities in a two-phase mixing layer have been identified. The first is related to the pinch between the shear branch and its surface-tension counterpart \cite{Otto_2013a}, while the second type arises from the pinch between the shear instability branch and the confinement-controlled branch \cite{Matas_2015b}. The dominant wave frequency and propagation speed for the surface-tension and confinement absolute instabilities are different: while the frequency for the former is higher, the wave propagation speed for the former is significantly lower than the latter. It is argued that the dominant mechanism for confinement absolute instability is inviscid in nature, since the interfacial wave propagation speed follows the Dimotakis speed \cite{Dimotakis_1986a}. The frequency for the surface-tension absolute instability has been found to be significantly higher than the interfacial wave frequency measured in experiment \cite{Otto_2013a}, and the lower frequency for the confinement absolute instability seemed to agree better with experimental data \cite{Matas_2015b}.  Nevertheless,  confinement absolute instability can only be observed when there is no velocity deficit in the velocity profile \cite{Matas_2015b}. The velocity deficit is a natural outcome of the wake of the separator plate and will gradually diminish away from the separator plate \tcr{\cite{Della-Pia_2024a}}. The discrepancy between the experiment and linear stability theory predictions is still not fully understood. 
\tcr{
For surface-tension absolute instability, both theoretical and experimental studies have identified that the dynamic pressure ratio (or momentum flux ratio) plays an essential role in the C/A transition \cite{Matas_2011a, Otto_2013a, Fuster_2013a, Della-Pia_2024a}. Since the gas-to-liquid dynamic pressure ratio does not involve fluid viscosity or surface tension, it remains unclear whether other parameters contribute to the C/A transition. Otto \etal \cite{Otto_2013a} showed that the critical dynamic pressure ratio for the C/A transition decreases with increasing surface tension, but a detailed study on the viscous effect, characterized, for example, by the Reynolds number, on the C/A transition is still lacking.
}

Detailed numerical simulation (DNS) has also been used to investigate the interfacial instability in two-phase mixing layers \cite{Fuster_2013a, Agbaglah_2017a, Ling_2017a, Ling_2019a, Bozonnet_2022a}. The advantage of DNS is that it allows for precise control of inflow conditions, making direct comparison with LSA easier. For certain ranges of parameters, previous studies showed that the dominant frequencies for the interfacial height measured in DNS agree well with LSA prediction for the surface-tension absolute instability \cite{Fuster_2013a}. However, the wave propagation speed observed follows the Dimotakis speed, which is significantly higher than the celerity predicted by LSA \cite{Jiang_2021a}. The physical reason behind the discrepancy remains unclear, leaving an important open question, i.e., is viscous effect indeed important to the interfacial stability and, if yes, to which features? To address these questions, parametric 2D interface-resolved simulations and linear stability analysis will be performed in the present study by varying the gas viscosity in a wide range. The goal of the present study is to characterize the effect of gas viscosity on the interfacial instability development and features, including C/A transition, wave speed, dominant frequency, and others.

The present study on the gas viscosity effect is also motivated by experimental and numerical studies on inlet gas turbulence modulation of interfacial instability \cite{Matas_2015a, Jiang_2020a, Jiang_2021a}. It was observed that when the interfacial instability lies in the absolute regime, the dominant frequency increases with the inlet turbulence intensity \cite{Jiang_2020a}. While the longitudinal shear instability varies, the secondary transverse instability and the spray formation downstream are also impacted \cite{Jiang_2021a}. Since inlet gas turbulence enhances the momentum transport, this effect can be approximately represented by a simple turbulent eddy viscosity model \cite{Pope_2000a}. By increasing the effective gas viscosity (sum of the gas dynamic and eddy viscosities), the linear stability analysis recovers the increasing trend of dominant frequency with turbulence intensity \cite{Matas_2015a, Jiang_2020a}. Nevertheless, discrepancies exist in the values predicted by stability analysis and DNS. Previous simulation results by Jiang and Ling \cite{Jiang_2021a} seemed to show that an increase in effective viscosity seems to cause the instability to transition to the absolute regime, though a more detailed investigation is required to confirm this point.

Since 3D interface-resolved simulations are too computationally costly for a parametric study to be completed in this study, 2D simulations will be used. 
\tcr{
Some 3D flow features in a two-phase mixing layer, such as turbulent flows \cite{Ling_2019a}, interfacial wave breakup \cite{Agbaglah_2021a}, and transverse interfacial instability \cite{Jiang_2021a}, will not be captured in the present simulations. Nevertheless, previous studies have demonstrated that 2D simulations are still capable of capturing longitudinal shear instability \cite{Fuster_2013a, Bozonnet_2022a} when inlet turbulence is absent.
} 
Though the focus is on the effect of gas viscosity, we will also vary the gas and liquid stream heights at the inlet and gas density to investigate the effects of confinement and density ratio on the influence due to varying gas viscosity. The rest of the paper is organized as follows: The governing equations and numerical methods for interface-resolved simulations will be presented in Section \ref{sec:sim}. The formulation for spatial-temporal viscous linear stability analysis and the results will be shown in Section \ref{sec:LSA}. The simulation results are presented and compared with the LSA predictions in Section \ref{sec:results}. Finally, conclusions will be drawn in Section \ref{sec:conclusions}.

\section{Simulation methods}
\label{sec:sim}
\subsection{Governing equations}
The liquid-gas two-phase flow is resolved using the one-fluid approach, wherein the two phases, liquid and gas, are treated as one fluid with material properties that change abruptly across the gas-liquid interface. The Navier-Stokes equations for incompressible flow with surface tension are given as 
\begin{equation}
  \rho (\partial_t \mathbf{u} + \mathbf{u} \cdot \nabla \mathbf{u}) = -\nabla p + \nabla \cdot (2 \mu \mathbf{D}) + \sigma \kappa \delta_s \mathbf{n},
  \label{eq:NS1}
\end{equation}
\begin{equation}
  \nabla \cdot \mathbf{u} = 0,
  \label{eq:NS2}
\end{equation}
 where $\rho$, $\mathbf{u}$, $p$, and $\mu$, represent density, velocity, pressure, and viscosity, respectively. The strain-rate tensor is denoted by $\mathbf{D}$. The surface tension term on the right-hand side of Eq.~\eqr{NS1} is a singular term, with the Dirac distribution function $\delta_s$ localized on the interface. The surface tension coefficient is represented by $\sigma$, while $\kappa$ and $\mathbf{n}$ are the local curvature and unit normal of the interface, respectively.
 
The two different phases are distinguished by the liquid volume fraction $C$, and $C=0$ and 1 indicate that the cells that are full of gas and liquid, respectively. For cells with interfaces, $0 < C < 1$. The evolution of $C$ satisfies the advection equation, 
\begin{equation}
  \partial_t C + \mathbf{u} \cdot \nabla C = 0.
  \label{eq:adv}
\end{equation}

The fluid density $\rho$ and viscosity $\mu$ are determined by 
\begin{align}
  \rho & = C \rho_l + (1 - C) \rho_g\,,
  \label{eq:density1}\\
  \mu &= C \mu_l + (1 - C) \mu_g\, ,
  \label{eq:viscosity1}
\end{align}
where the subscripts $g$ and $l$ correspond to the gas and the liquid phases, respectively.

\subsection{Numerical methods}
The governing equations are solved by the finite volume method on a staggered grid. The advection equation, Eq.~\eqr{adv}, is solved using a geometric volume-of-fluid (VOF) method. The interface normal is computed  following the mixed Young's-centred method \cite{Aulisa_2007a}. The Lagrangian-explicit scheme \cite{Li_1995b} is used for the VOF advection \cite{Scardovelli_2003a}.  The convection term in the momentum equation, Eq.~\eqr{NS1}, is discretized consistently with the VOF method \cite{Arrufat_2020a}, and this mass-momentum consistence has been shown to be crucial in capturing interfacial dynamics when large velocity and density contrasts are present at the interface \cite{Rudman_1998a, Ling_2017a,Vaudor_2017a, Zhang_2020a, Arrufat_2020a}. The incompressibility condition is incorporated using the projection method \cite{Chorin_1968a}. The pressure Poisson equation is solved using PFMG multigrid solver in the HYPRE library. The viscous term is discretized explicitly using the second-order centered difference scheme. The interface curvature is calculated using the height-function method \cite{Popinet_2009a} and the balanced continuous-surface-force method is used to discretize the surface tension term  \cite{Renardy_2002a, Francois_2006a, Popinet_2009a}. The time integration is done by a second-order predictor-corrector method.  The above numerical methods have been implemented in the open-source solver, \emph{PARIS-Simulator}. Detailed implementations and validation of the code can be found in previous studies \cite{Tryggvason_2011a, Ling_2015a, Ling_2017a, Ling_2019a, Arrufat_2020a, Aniszewski_2021a}.

\subsection{Problem description and key dimensionless parameters}
\label{sec:parameters}
\begin{figure}[tbp]
  \centering
  \includegraphics[width=0.99\textwidth, trim = {2.4cm 6.5cm 2.4cm 0.5cm},clip]{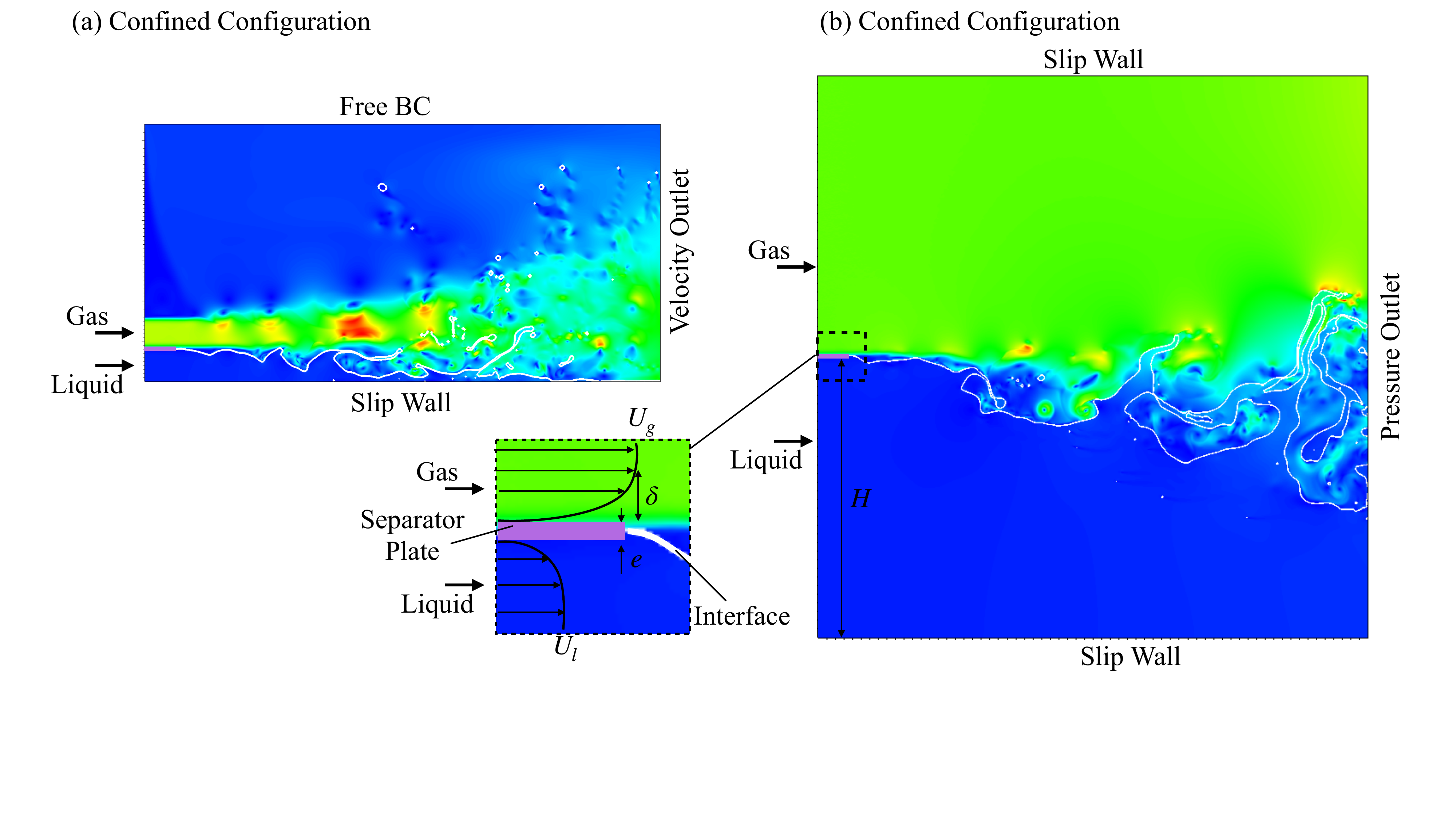}
\caption{\tcr{Simulation setups for (a) the confined configuration with small stream height ($\eta=8$) and (b) the unconfined configuration with large stream height ($\eta=65$).}}
\label{fig:Computational_domains} 
\end{figure}

Two different configurations of two-phase mixing layers are considered in the present study, see Fig.~\ref{fig:Computational_domains}. In the first configuration, see Fig.~\ref{fig:Computational_domains}(a), two parallel liquid and gas streams enter the rectangular domain from the left. The heights of the liquid and gas streams are small compared to the domain height. In contrast, the gas and liquid stream heights are enlarged to half of the domain height in the second configuration, and the domain height is also larger, see Fig.~\ref{fig:Computational_domains}(b). The purpose of increasing the stream and domain heights is to elevate the confinement effect. The boundary conditions for the two configurations are similar, which will be discussed later. 

The liquid density $\rho_l$ and viscosity $\mu_l$ are similar to water and are kept constant, while the gas density $\rho_g$ and viscosity $\mu_g$ are varied. At the inlet, the two streams are separated by a separator plate, \tcr{see Fig.~\ref{fig:Computational_domains}}. The streams meet at the end of the separator plate. The separator plate thickness $e$ is taken to be $\delta/4$, where $\delta$ is the gas boundary layer thickness. According to the previous study by Fuster \etal  \cite{Fuster_2013a}, the effect of $e$ on the interfacial instability becomes negligible when it is significantly smaller than $\delta$. The gas and liquid velocities away from the separator plate are uniform and represented by $U_g$ and $U_l$, respectively. The fluid properties and inflow conditions are are given in Table~\ref{tab:phy_para1}, chosen to be similar to the previous studies \cite{Ling_2017a, Ling_2019a, Jiang_2021a}, except for the variation of $\mu_g$.

\begin{table}[tbp]
  \begin{center}
  \begin{tabular}{ccccccccccc} 
  \hline
  Series&    $\rho_{l} $&$\rho_{g} $&$ \mu_{l} $  &$ \mu_{g} $ &$\sigma$ &$H$ &$U_g$ &$U_l$ &$e$ &$\delta$\\   
&   (kg/m$^{3}$) &  &(Pa$\cdot$s) &($10^{-5}$Pa$\cdot$s) &(N/m) &(mm) &(m/s) & &(mm) & \\
   \hline
  A &  	  &50 &					& 5\ - 89.5 & 		&0.8 & 	& 	& 		&\\ 
  B & 1,000 &50 &$\mathrm{10^{-3}}$ 	& 5\ - 89.5 &0.05 	&6.5 &10 	&0.5 &0.025 	&0.1\\ 
  C & 	  &12.5 &					& {0.6\ - 30} & 		&0.8 &	&	& 		&\\ 
  \hline
  \end{tabular}
  \caption{Physical properties \& geometric parameters for all simulation cases.}  
   \label{tab:phy_para1}
    \end{center}
\end{table}

There are two physical length scales involved in the present problem: the gas boundary layer thickness, $\delta$, and the gas stream height at the inlet, $H$. Consistent with previous studies, we have considered the liquid boundary layer thickness and the liquid stream height to be similar to their gas counterparts, namely $\delta_g=\delta_l=\delta$  and $H_g=H_l=H$ \cite{Matas_2011a, Fuster_2013a, Ling_2017a, Ling_2019a, Bozonnet_2022a}. \tcr{For convenience, we will refer to $\delta$ and $H$ as boundary layer thickness and stream height in the rest of the paper, since their values for gas and liqui are the same.} The ratio between $H$ and $\delta$, i.e., $\eta=H/\delta$, characterizes the effect of confinement due to the finite stream thickness \cite{Bozonnet_2022a}. When $\eta \gg 1$, as in the second configuration, $\delta$ is the only relevant length scale, which is  often used to define the Reynolds and Weber numbers, $\text{Re}={\rho_gU_g\delta}/{\mu_g}$ and $\text{We}={\rho_gU_g^2\delta}/{\sigma}$ \cite{Otto_2013a, Fuster_2013a}. However, when $\eta$ is not sufficiently large, the confinement from the stream boundary will influence the interfacial instability \cite{Matas_2015b, Bozonnet_2022a}. 
\tcr{
Two different values of $H$ (or $\eta$) are considered: the confined configuration with $\eta=8$ and the unconfined configuration with $\eta=64$, as shown in Figs.~\ref{fig:Computational_domains}(a) and (b), respectively. In the present study, simulation series A and B have similar parameters, except that series A is for the confined configuration, while series B is for the unconfined one. The difference between the results from these two series can then be used to characterize the effect of confinement. Furthermore, simulation series C uses the same confined configuration as series A but with a reduced $\rho_g$. Therefore, comparing the results from series A and C serves to characterize the effect of the density ratio $r$.
}

\tcr{
A wide range of $\mu_g$ is considered in each simulation series. For simulation series A and B with density ratio $r=0.05$, $\mu_g$ varies from $5 \times 10^{-5}$ to $8.95 \times 10^{-4}$ Pa-s. Correspondingly, the viscosity ratio $m=\mu_g/\mu_l$ increases from 0.006 to 0.9, and the Reynolds number $\mathrm{Re}=\rho_g U_g \delta/\mu_g$ decreases from 1,000 to 56, as shown in Table~\ref{tab:cases}. For simulation series C, where $r=0.0125$, $\mu_g$ varies from $6 \times 10^{-6}$ to $3 \times 10^{-4}$ Pa-s. As a result, $m$ and $\mathrm{Re}$ vary from 0.006 to 0.3 and 2000 to 42, respectively. The ranges of $\mu_g$ and $\text{Re}$ are chosen to be broad enough to capture both the convective and absolute regimes, with a sufficiently large number of cases (about 10 for each series) run to identify the critical Reynolds number ($\text{Re}_{cr}$) at which the A/C transition occurs. In total, approximately 30 simulations are performed.
}

\begin{table}[tbp]
  \begin{center}
  \begin{tabular}{cccccccc}
  \hline
  Series &M &Re &We &r &m &$\eta$ &\tcr{Configuration} \\
 &   $\frac{\rho_gU_g^2}{\rho_lU_l^2}$ & $\frac{\rho_gU_g\delta}{\mu_g}$ &$\frac{\rho_gU_g^2\delta}{\sigma}$ & $\frac{\rho_g}{\rho_l}$ & $\frac{\mu_g}{\mu_l}$ &$\frac{H}{\delta}$ \\
 \hline
  A &20 &$1,000 - 56$ &10 &0.05 &$0.05 - 0.9$ &8 &\tcr{Confined}  \\
  B &20 &$1,000 - 56$ &10 &0.05 &$0.05 - 0.9$ &65 &\tcr{Unconfined} \\
  C &5 &$2000 - 42$ &2.5 &0.0125 &$0.00625 - 0.3$ &8 &\tcr{Confined} \\
  \hline
  \end{tabular}
\caption{Dimensionless parameters for all simulation cases. \tcr{Simulation series A and C are for the confined configuration with a small stream height (see Fig.~\ref{fig:Computational_domains}(a)), and series B is for the unconfined configuration with a large stream height (see Fig.~\ref{fig:Computational_domains}(b)).} }
\label{tab:cases}
\end{center}
\end{table}

\subsection{Simulation setup}
The computational domains and boundary conditions for the two configurations are depicted in Fig.~\ref{fig:Computational_domains}. The BC and domain size in the first configuration (Fig.~\ref{fig:Computational_domains}(a)) are identical with the previous studies \cite{Ling_2017a, Ling_2019a, Jiang_2021a}. The length and height of the domain are $L_x=16H$ and $L_y=8H$, respectively. The domain is initially filled with stationary gas. A Dirichlet boundary condition is specified for the velocity at the left surface. While the tangential velocity components are $v=w=0$, the normal component is
\begin{equation}
	u(y) = 
	\left \{
		\begin{array}{ll}
		(U_{l}+U'_l)\,\mathrm{erf} \left[\frac{H-y}{\delta}\right], &  -H \le y < 0, \\
		0,                                               &  0 \leq y < e, \\
		U_{g}\,\mathrm{erf}\left[\frac{y-(H+e)}{\delta}\right]\mathrm{erf}\left[\frac{2H-y}{\delta}\right], & e \leq y < H, \\
		0, &  \mathrm{else}. 
		\end{array} 
	\right.
	\label{eq:vel_BC1}
\end{equation}
The velocities in the gas and liquid streams away from the separator plates ($0 \leq y < e$) are uniform, equating to $U_l$ and $U_g$, respectively. The error function, $\mathrm{erf}$, is employed to model the velocity profile in the boundary layers adjacent to the separator plates. Small-amplitude perturbations are introduced to the inlet liquid velocity, which can be expressed as the sum of $N$ sinusoidal functions with different frequencies,
\begin{equation}
	\frac{U'_l}{U_l} =\xi^*_{pert} \left[ \frac{1}{N} \sum_{k=1}^N \sin(2\pi f_{pert,k} t + \phi_k) \right],
	\label{eq:pert_sine}
\end{equation}
where $f_{pert,k}$ and $\phi_k$ are the frequency and phase of the $k^{th}$ perturbation mode and the perturbation amplitude $\xi^*_{pert}$ is the normalized perturbation amplitude. Both single-mode ($N=1$) and multi-mode ($N=10$) perturbations are considered. In the first configuration, the velocity outflow boundary condition is specified on the right boundary. The bottom surface is a slip wall, while the top is a free boundary, thorough which allows fluid to freely enter or leave the domain. Verification of the boundary conditions and domain size can be found in previous study \cite{Ling_2019a}. 

In the second configuration shown in Fig.~\ref{fig:Computational_domains}(b), the gas and liquid streams are extended to the top and bottom of the domain. The inlet velocity boundary condition is adjusted accordingly as 
\begin{equation}
	u(y) = 
	\left \{
		\begin{array}{ll}
		(U_{l}+U'_l)\,\mathrm{erf} \left[\frac{H-y}{\delta}\right], &  y < 0, \\
		0,                                               &  0 \leq y < e, \\
		U_{g}\,\mathrm{erf}\left[\frac{y-(H+e)}{\delta}\right], &  y> e . 		
		\end{array} 
	\right.
	\label{eq:vel_BC2}
\end{equation}
A large stream thickness $H=65\delta$ is used. Additional tests were made confirm that $\eta=65$ is enough to elevate the confinement effect and the detail value of $H$ is immaterial. Both the top and bottom boundaries are taken as slip walls. Different from the first configuration, the pressure outflow boundary condition is specified on the right surface. Similar boundary conditions were used in the previous studies by  Agbaglah \etal \cite{Agbaglah_2017a} and Bozonnet \etal \cite{Bozonnet_2022a}. 

A uniform Cartesian mesh is used to discretize the domain. The cell size $\Delta x=\Delta y =H/256=6.25$ \textmu m is shown to be sufficient to resolve the interfacial instability under similar parameter ranges. Additional grid-refinement studies are performed, and the results are shown in \ref{sec:grid_refinement}, confirming that the current mesh resolution yields mesh-independent results. 

By selecting $U_g$ and $\delta$ as the velocity and length scales, the dimensionless variables, denoted by superscript $^*$, are defined as 
\begin{align}
  t^* = \frac{tU_g}{\delta},\  x^* = \frac{x}{\delta},\  u^* = \frac{u}{U_g}. 
  \label{eq:x_star}
\end{align}
All simulation cases were run to at least $t^*=6000$. This long time duration ensures statistically converged results and high-resolution frequency spectra. Simulation results for various simulation run times are displayed in \ref{sec:sim_time_study}, to verify the simulation time is sufficiently long to obtain accurate spectra. 

\section{Linear stability analysis \tcr{for unconfined configuration}}
\label{sec:LSA}
\subsection{Orr-Sommerfeld equations and numerical methods}
Viscous linear stability analysis (LSA) is performed to better understand the effect of gas viscosity on interfacial instability. A spatio-temporal analysis is conducted. \tcr{For LSA, the unconfined configuration is considered and the stream height is significantly larger than the inlet boundary layer thickness, and thus the value $\eta$ is immaterial to the results. In LSA, different $\eta$ were tested, the resutls showed that $\eta=15$, which was also used in the previous study \cite{Otto_2013a}, is generally enough to yield results that are independent of $\eta$ and the confinement effect.} The velocity profiles for parallel base liquid and gas flows, $u_{l,0}(y)$ and $u_{g,0}(y)$, are specified following previous studies \cite{Otto_2013a, Fuster_2013a, Ling_2019a}, which is similar to velocity boundary conditions (Eq.~\eqref{eq:vel_BC2}) used in the simulation \tcr{for the unconfined configuration}, 
\begin{align}
		u_{l,0}(y)  = -U_l\mathrm{erf}(\frac{y}{\delta}) & +U_0[1+\mathrm{erf}(\frac{y}{\delta_d})], \quad  y < 0, \label{eq:vel_liq_bases}\\
		u_{g,0}(y) = \ \  U_g\mathrm{erf}(\frac{y}{\delta}) & +U_0[1-\mathrm{erf}(\frac{y}{\delta_d})], \quad y> 0.  \label{eq:vel_gas_bases}
\end{align}
\tcr{This velocity profile model is found to agrees very well with the mean flow data measured in recent experiment of two-phase mixing layers \cite{Della-Pia_2024a}.}
Here, the interface is located at $y=0$. Similar to the simulations, the boundary layers of the gas and liquid streams are modeled using error functions, and the thicknesses of which are the same by $\delta_g=\delta_l=\delta$. 

It is suggested by Otto \etal \cite{Otto_2013a} that the interfacial velocity, $U_0$ can be specified as 
\begin{equation}
U_0=\frac{\left(\frac{\mu_gU_g}{\delta}+\frac{\mu_lU_l}{\delta} \right) \delta_d}{\mu_g+\mu_l}. 
\label{eq:interface}
\end{equation}
which will guarantee stress continuity at the interface. The effect of the separator plate on the streamwise velocity profile, mainly the velocity deficit created by the wake, is modeled by the correction functions $U_0[1 \pm \mathrm{erf}(y/\delta_d)]$, where $\delta_d$ is the parameter to control $U_0$ and the deficit in the velocity profile. Since the velocity profile varies in the streamwise directions, the selection of $\delta_d$ and $U_0$ is somewhat subjective, and different values have been used in the previous studies. In the present study, we have used $\delta_d=0$, which will give $U_0=0$. The model velocity profile corresponds to the exit of the separator plate where the gas and liquid stream just met. If a finite $\delta_d$ or $U_0$ was used, the velocity profile would be different for different $\mu_g$, in other words, the base state will change. In the present analysis, we have excluded the viscous effect on the base velocity profile and solely focused on the effect of gas viscosity on the perturbations. A systematic parametric study for effects of $\delta_d$ and $U_0$ will be relegated to our future work. It is worth noting that the key finding of the present study remains valid for finite $U_0$ as will be discussed later and shown in \ref{sec:effect_infc_vel}. 

The normal-mode perturbation is introduced in the form of stream function, 
\begin{equation}
\psi_j(x,y,t) = \varphi_j(y) e^{i(\alpha x - \omega t)}, 
\label{eq:pert}
\end{equation}
where $\alpha$ and $\omega$ are the complex wavenumber and frequency, respectively. Substituting the perturbation in the linearized Navier-Stokes equation, it yields the Orr-Sommerfeld equations, the dimensionless form of which using the dimensionless variables defined in Eq.~\eqref{eq:x_star} are given as 
\begin{align}
&\left[(-i\omega^* + i\alpha^* u_{l,0}^*)\Big(\pds{}{(y^*)} - (\alpha^*)^2\Big) - \frac{r}{m \mathrm{Re}}\Big(\pds{}{(y^*)} - (\alpha^*)^2\Big)^2 - i\alpha^* \ods{u_{l,0}^*}{(y^*)}\right]\varphi^*_l = 0, \quad  y < 0, \nonumber \\ 
&\left[(-i\omega^* + i\alpha^* u_{g,0}^*)\Big(\pds{}{(y^*)}  - (\alpha^*)^2\Big) - \quad \frac{1}{\mathrm{Re}}\Big(\pds{}{(y^*)} - (\alpha^*)^2\Big)^2 - i\alpha^* \ods{u_{g,0}^*}{(y^*)}\right]\varphi_g^* = 0\, , \quad  y >0, 
\label{eq:Orr_g}
\end{align}
which can be solved along with the boundary conditions for velocity and stresses at the interface. For temporal analysis, Eq.~\eqref{eq:Orr_g} is solved for complex frequencies $\omega^* = \omega_r^*+i\omega_i^*$ when a real wavenumber $\alpha^*$ is given. In contrast, for spatial analysis, the complex wavenumber $\alpha^* = \alpha_r^*+i\alpha_i^*$ is determined for a given real frequency $\omega^*$. In the current spatio-temporal analysis, both the frequency and wavenumber are complex numbers, and the complex wavenumbers $(\alpha_r^*, \alpha_i^*)$ are solved given complex frequencies $(\omega_r^*, \omega_i^*)$. 

The Orr-Sommerfeld equations are discretized by a Chebyshev collocation method, see the previous study \cite{Otto_2013a} for details. For both phases, 150 Chebyshev polynomials are used, which have been tested to be sufficient to yield converged results. 

\subsection{Effect of Re on C/A transition}
\begin{figure}[tbp]
 \centering
  \includegraphics[width=1.\textwidth, trim = {3.5cm 10cm 4.5cm 5.5cm},clip]{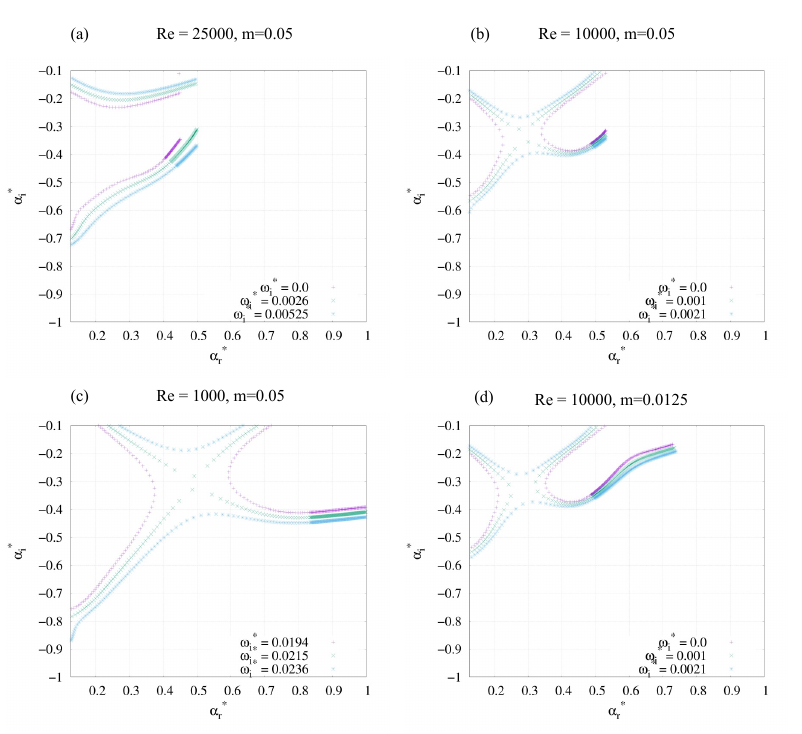}
  \caption{Spatial branches for different $\text{Re}$ and $m$. For all cases, $r = 0.05, M = 20$ and $\delta_d = 0.0$.}
\label{fig:Saddle_point_m_Re_effect} 
\end{figure}

The spatial branches for the cases at different $\text{Re}$ and $m$ are shown in Fig.~\ref{fig:Saddle_point_m_Re_effect}. Other parameters, including $r=0.05$, $M=20$, and $\text{We}=20$, are fixed. For a sufficiently large positive $\omega_i^*$, two distinct spatial branches appear on the $\alpha_r^*$-$\alpha_i^*$ plane, as seen in Fig.~\ref{fig:Saddle_point_m_Re_effect}(a). The points on a branch correspond to different $\omega_r^*$. According to the principle of causality, the upper and lower branches represent downstream and upstream propagating perturbations, respectively. As $\omega_i^*$ decreases, the branches move: typically, the upper branches move down, while the lower one moves up. For the case with $\mathrm{Re}=25000$, $m=0.05$, the upper branch crosses the imaginary axis ($\alpha_i=0$) before $\omega_i^*$ reaches zero. This indicates that there exists a certain range of $\alpha_r^*$ that is convectively unstable ($\alpha_i<0$).

When $\text{Re}$ decreases from 25,000 to 10,000, it can be observed that the upper branch for the same $\omega_i^*$ moves downward, while the lower counterpart moves up. Eventually the upper and lower branches pinch, creating a saddle point, before $\omega_i^*$ reaches zero, as depicted in Fig.~\ref{fig:Saddle_point_m_Re_effect}(b). The appearance of the saddling point indicates that the convective instability transitions to absolute instability \cite{Briggs_1964a, Huerre_1990a}. Conventionally, the transition from absolute to convective regimes (A/C) is often characterized by the dynamic pressure ratio $M$ \cite{Fuster_2013a} or the velocity difference $(U_g-U_l)/(U_g+U_l)$ \cite{Otto_2013a}. The present results indicate that, with $r$, $\eta$, and $M$ held constant, an A/C transition can be triggered by just increasing the Reynolds number. This finding is supported by numerical simulation results, which will be presented later.

Here, the variation of $\text{Re}$ is introduced by varying the gas viscosity. Yet, when the gas viscosity decreases, both $\text{Re}$ increases and $m$ decreases simultaneously. In order to investigate their individual effects on the A/C transition, additional cases for varying $\text{Re}$ with fixed $m$ and for varying $m$ with fixed $\text{Re}$ are studied. When $\text{Re}$ increases from 10,000 to 25,000 with $m=0.05$ fixed, the instability transitions from absolute to convective, as seen in Figs.~\ref{fig:Saddle_point_m_Re_effect}(a) and (b). In contrast, when $m$ is reduced from 0.05 to 0.0125, while keeping $\text{Re} = 10,000$ constant, the instability remains absolute, as seen in Figs.~\ref{fig:Saddle_point_m_Re_effect}(b) and (d). The results thus indicate that $\text{Re}$ plays a more significant role than $m$, and the increase of $\text{Re}$ is the dominant reason for the A/C transition.

\subsection{Effect of Re on the saddle-point features}
\begin{figure}[tbp]
 \centering
  \includegraphics[width=1.\textwidth, trim = {3.5cm 8.5cm 3.5cm 7.5cm},clip]{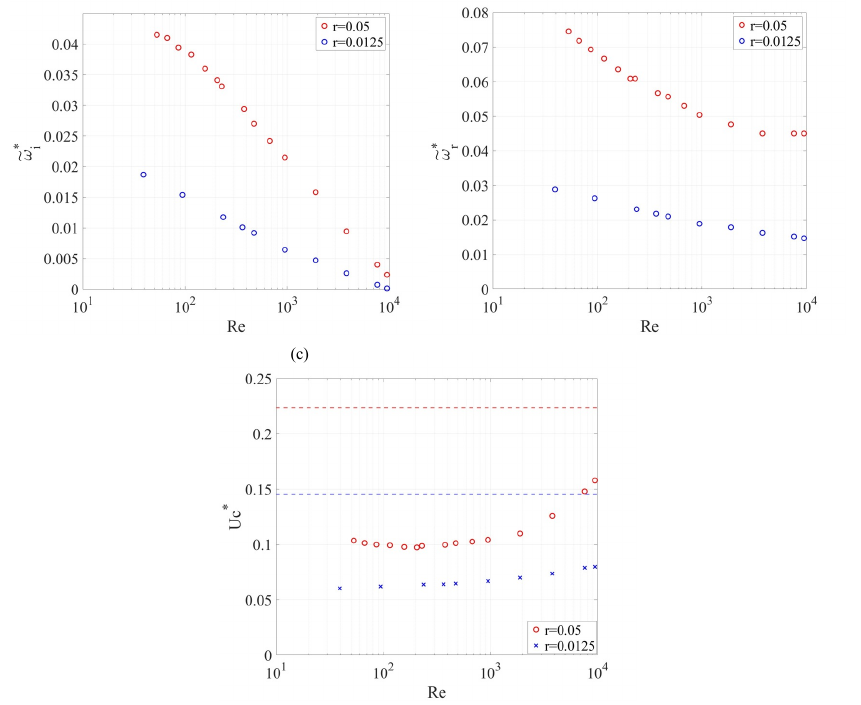}
\caption{Variations of saddle-point features with $\text{Re}$: (a) temporal growth rate $\tilde{\omega}_i^*$, (b) frequency $\tilde{\omega}_r^*$, and (c) celerity $U_c^*$. For all cases shown, $M = 20$, $m = 0.05$, and $U_0 = 0.0$. The dashed lines in (c) represent the Dimotakis speed $U_D^*$ for the corresponding density ratio $r$.}
  \label{fig:sp_properties_Re} 
\end{figure}

Although both the cases $\text{Re}=10,000$ and 1,000 with $m=0.05$ exhibit a saddle point and thus are in the absolute instability regime, the features of the saddle point vary as $\text{Re}$ changes. It is observed from Figs.~\ref{fig:Saddle_point_m_Re_effect}(b) and (c) that the position of the saddle point moves to the right, when $\text{Re}$ decreases from 10,000, to 1,000, indicating an increase in the wavenumber for the most unstable mode, $\tilde{\alpha}_r^*$, where $\tilde{()}$ is used to represent saddle-point features. Furthermore, the temporal growth rate for the saddle point, $\tilde{\omega}_i^*$, for $\text{Re}=10,000$ is much smaller than that for $\text{Re}=1000$. The variation of $\tilde{\omega}_i^*$ as a function of $\text{Re}$ is shown in Fig.~\ref{fig:sp_properties_Re}(a). Generally, $\tilde{\omega}_i^*$ decreases monotonically with $\text{Re}$. When $\tilde{\omega}_i^*$ reaches zero, the absolute instability transitions to its convective counterpart. The critical value for transition to occur is located around $\text{Re}=10,000$. The present LSA analysis assumes the interfacial velocity $U_0=0$ ($\delta_d=0$). Additional case with $U_0=0.1U_l$ has been studied and the results are shown in \ref{sec:effect_infc_vel}. It is shown that the results are quite similar for different $U_0$ and the critical Reynolds number $\text{Re}_{cr}$ is not sensitive to the value of $U_0$. 

Moreover, when an absolute mode exhibits a small $\tilde{\omega}_i^*$, it may not dominate the concurrent convective modes. Thus, we refer to this transition regime between the typical absolute and convective regimes as the \emph{weak absolute instability} regime, although the boundary between absolute and weak absolute regimes is somewhat blurry. The weak absolute instability regime can also be recognized from the numerical simulation results, which will show other interesting features for this regime.

The variations of the frequency $\tilde{\omega}_r^*$ and celerity $U_c^* = \tilde{\omega}_r^*/\tilde{\alpha}_r^*$ at the saddle point are shown in Figs.~\ref{fig:sp_properties_Re}(b) and (c), respectively. Similar to the temporal growth rate, the frequency $\tilde{\omega}_r^*$ generally decreases with $\text{Re}$. Although experimental and numerical studies varying the gas viscosity are lacking in the literature, experiments and simulations adjusting the inlet gas turbulence intensity \cite{Matas_2015a, Jiang_2020a, Jiang_2021a} have indicated that the frequency for the absolute mode decreases when the effective eddy viscosity increases, which are consistent with the present results. 

The variation of the celerity $U_c^*$ is more complex. For the density ratio $r=0.05$, a non-monotonic variation is observed, \ie, $U_c^*$ first decreases and then increases as $\text{Re}$ increases. The values of $U_c^*$ are significantly smaller than the normalized Dimotakis speed, $U_D^*=U_D/U_g$, with the same density ratio $r$. The Dimotakis speed $U_D$, expressed as 
\begin{equation}
U_D = \frac{\sqrt{\rho_g}U_g+\sqrt{\rho_l}U_l}{\sqrt{\rho_g}+\sqrt{\rho_l}}\,,
\label{eq:Ud}
\end{equation}
is obtained based on the assumption that the gas and liquid dynamic pressures are in a balance in the reference frame moving with the wave speed \cite{Dimotakis_1986a}, and therefore is an inviscid feature. For the absolute mode determined by the present viscous stability analysis, it is expected that  the wave propagation speed will follow the celerity $U_c$. Former studies have shown that the wave propagation speed agrees with the Dimotakis speed $U_D$ as the wave amplitude grows \cite{Lasheras_2000a, Ling_2019a, Jiang_2021a}. The wave speed transition from $U_c$ to $U_D$ will be further examined in the next section.

\section{Simulation results}
\label{sec:results}

\tcr{
The numerical simulation results for both the confined and unconfined configurations, as shown in Figs.~\ref{fig:Computational_domains}(a) and (b) respectively, for the two-phase mixing layer will be presented. We will begin with the results for series A (confined configuration), discussing the general effect of gas viscosity in section \ref{sec:general}, its impact on interfacial instability in section \ref{sec:nearfield}, and the non-linear interactions between interfacial waves in section \ref{sec:farfield}. Next, we will present the results for series B (unconfined configuration) in section \ref{sec:confinement} and discuss the effect of confinement by comparing the results for series A and B, which share identical parameters except for stream height. Finally, we will present the results for series C, which features a lower gas-to-liquid density ratio, in section \ref{sec:densityratio}. The comparison between the results for series A and C in the confined configuration will illustrate the effect of the density ratio.
}

\subsection{General behaviour}
\label{sec:general}
\begin{figure}[tbp]
    \centering
    \includegraphics[width=0.99\textwidth, trim = {4.5cm 11.cm 5.cm 8.5cm},clip]{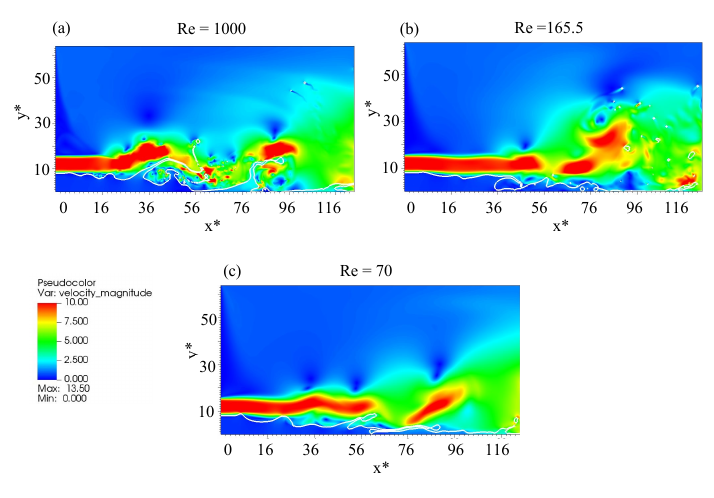}
\caption{Representative snapshots for the two-phase mixing layers with different $\text{Re}$ \tcr{in series A for the confined configuration, see Table \ref{tab:cases}}. The colored background indicates the velocity magnitude. The white solid lines delineate the gas-liquid interfaces.}
    \label{fig:Flow_behaviour_confined}
\end{figure}

\tcr{The simulation results for different $\text{Re}$ in series A (confined configuration) are presented in Fig.~\ref{fig:Flow_behaviour_confined} to qualitatively show the effect of gas viscosity on the interfacial instability and interfacial wave development.} Based on the LSA results given above, we mainly use $\text{Re}$ to represent cases with different $\mu_g$, since it plays a more important role than $m$. 

After the gas and liquid streams meet at the end of the separator plate, shear instability induces wavy structures on the interface. The interfacial waves propagate downstream, and their amplitudes grow in the streamwise direction. The interfacial dynamics near the inlet differ from those in the far field, where the wave amplitude becomes higher than the characteristic length $\delta$. For convenience of discussion of the wave dynamics, we loosely define the near and far fields of the two-phase mixing layer based on the interface wave amplitude, \ie, the near field is for the wave amplitude $\bar{h}\lesssim \delta$, while the far field is considered to be the spatial region where $\bar{h}\gtrsim \delta$. A detailed definition of the time-averaged wave amplitude $\bar{h}$ will be given later.

\paragraph{Near-field interfacial instability}
The focus on the near field is on the interfacial instability. Particular attention is paid on the effect of $\text{Re}$ on the frequency for the most unstable mode. The qualitative difference in the near fields for different $\text{Re}$ can be recognized in Fig. \ref{fig:Flow_behaviour_confined}. When $\text{Re}$ decreases, the wave amplitude grows more rapidly in space. Nevertheless, the important features including the dominant frequency variation cannot be directly observed in the snapshots and spectral analysis is required to demonstrate them. 

\paragraph{Far-field non-linear wave dynamics}
The study of the far field will be focused on the effect of $\text{Re}$ on non-linear wave dynamics. First of all, the modulation of interfacial instability in the near field will influence the spatial development and the subsequent breakup of interfacial waves in the far field. Furthermore, as $\text{Re}$ decreases, the gas stream becomes more stable. As a result, the gas velocity decreases more rapidly in space, which in turn weakens the interaction between the interfacial wave and the interfacial wave, as can be observed in Fig.~\ref{fig:Flow_behaviour_confined}. 

\subsection{Effect of gas viscosity on the near-field interfacial instability}
\label{sec:nearfield}
The results of simulation series A \tcr{(confined configuration)} are presented in this section to illustrate the effect of gas viscosity on the interfacial instability in the near field. 

\subsubsection{Charaterizing different instability regimes}
\begin{figure}[tbp]
    \centering
    \includegraphics[width=1\linewidth, trim = {1.1cm 10.5cm 1.3cm 5.5cm},clip]{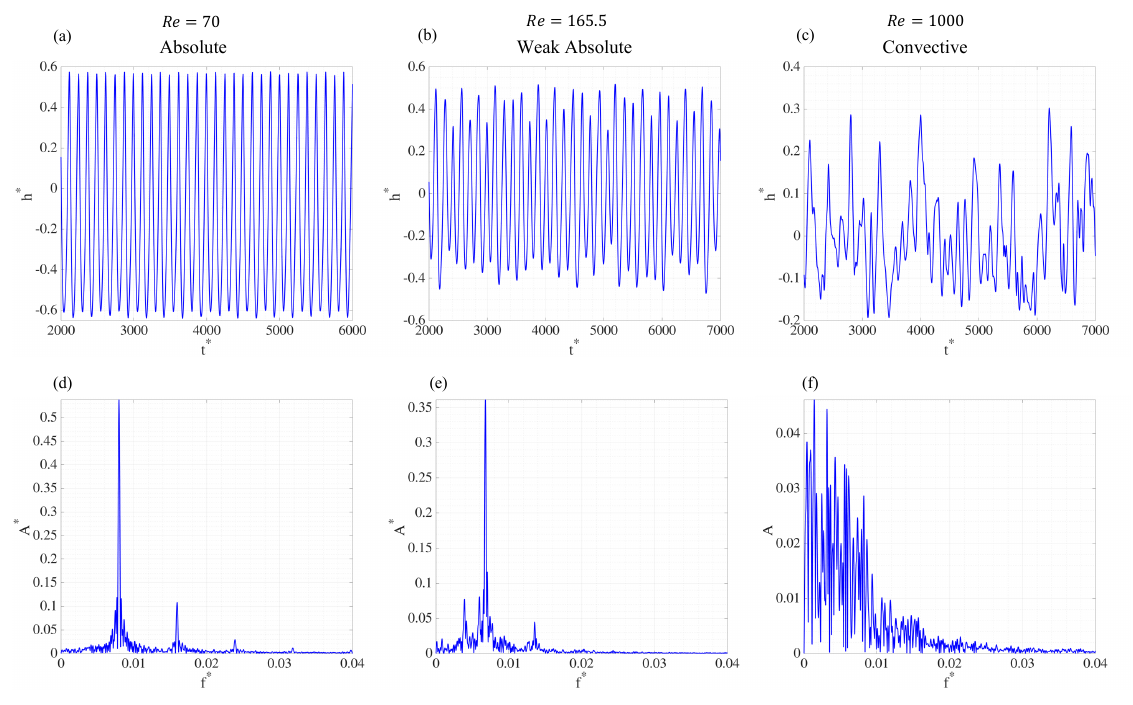}
	\caption{(a)-(c) Temporal evolutions and (d)-(f) frequency spectra of the interfacial height at \(x^* = 3\) for different $\text{Re}$, showcasing various instability regimes: (a, d) $\text{Re} = 70$ (absolute instability), (b, e) $\text{Re} = 165.5$ (weak absolute instability), and (c, f) $\text{Re} = 1000$ (convective instability). All cases are from series A.}
	\label{fig:Effect_Re_nopert}
\end{figure}

The temporal evolutions of the interfacial height $h$ for different $\text{Re}$ are shown in the first row of Fig.~\ref{fig:Effect_Re_nopert}, and the corresponding frequency spectra are shown in the second row.  The interfacial height $h$ is the vertical distance from the domain bottom to the interface, which depends on $x$ and $t$, \ie, $h(x,t)$. No perturbation is introduced at the inlet for the results here, namely $\xi^*_{pert}=0$.  The results shown are measured at  $x^* = 3$, which is close to the inlet.  It can be observed that the wave amplitudes remain significantly lower than $\delta$, \ie, $|h^*|<1$. Different instability regimes can be identified. When $\text{Re}=70$, a dominant frequency, \tcr{denoted as $f^*_0$}, appears in the spectrum, at which the interfacial height oscillates in time, see Figs.~\ref{fig:Effect_Re_nopert}(a,d). The other smaller spikes in the spectrum are located at frequencies that are integer times of the absolute frequency and are induced by the nonlinear effect. In contrast, for $\text{Re}=1000$, the spectrum is noisy, because the instability is convective and the dynamic system acts as a noise amplifier. Multiple convective modes grow and their amplitudes are sensitive to inlet conditions, see Figs.~\ref{fig:Effect_Re_nopert}(c,f). There is no frequency that dominate the whole space. For the intermediate $\text{Re}=165.5$, the instability is in the weak absolute regime, for which the spectrum shows both the dominant absolute mode but also the noises created by the convective modes, see Figs.~\ref{fig:Effect_Re_nopert}(b,e).

\begin{figure}[tbp]
  \centering
  \includegraphics[width=1.0\textwidth, trim = {2cm 3.5cm 2.2cm 4.9cm},clip]{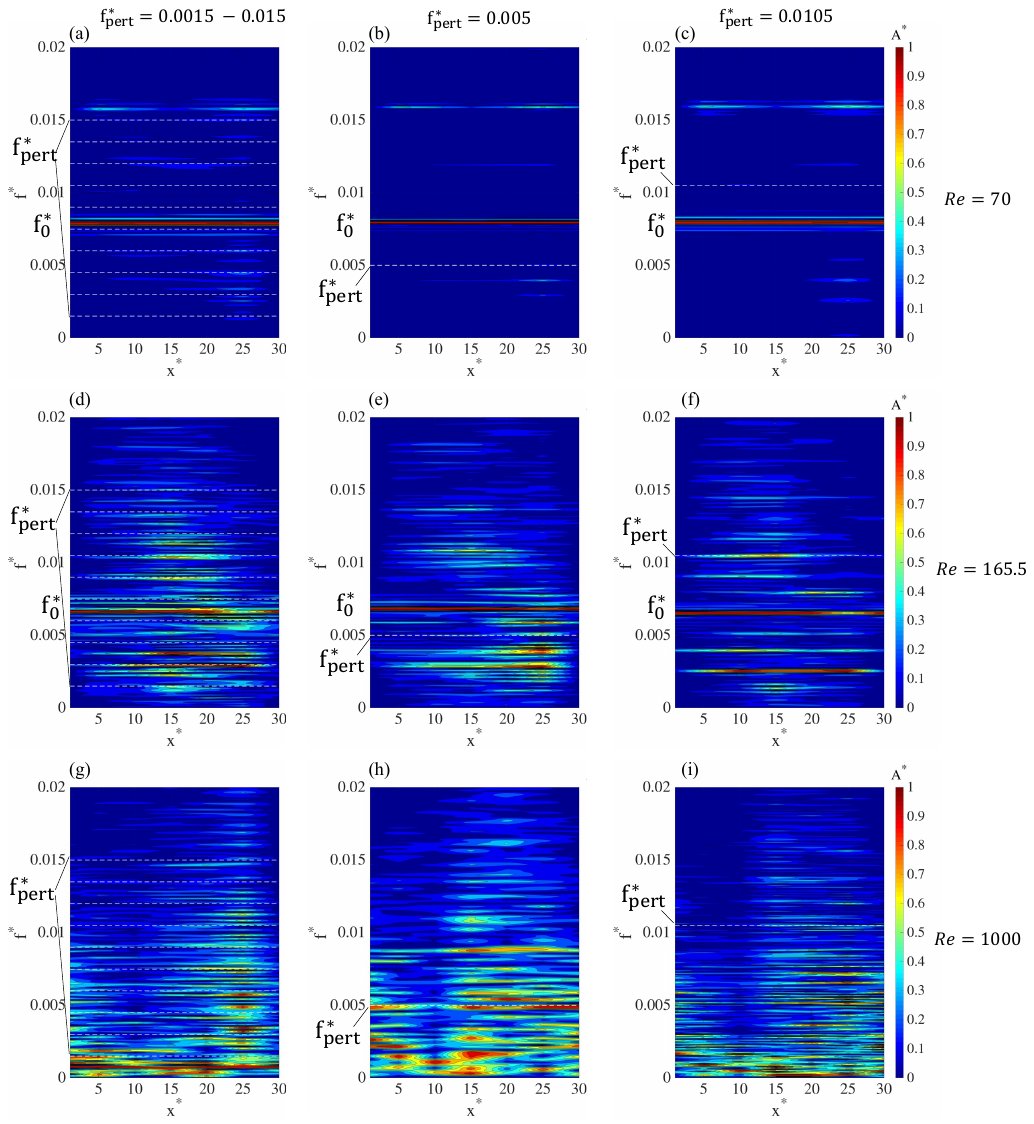}
	\caption{Spectrograms for different instability regimes. The first, second, and third rows correspond to $\text{Re}=70$ (absolute), 165.5 (weak absolute), and 1000 (convective), respectively. Different columns are for different inlet perturbation frequencies: the first column represents multi-mode perturbations with 10 different frequencies varying from $f^*_{pert}=$0.0015 to 0.015, the second and third columns are for single-mode perturbation with $f^*_{pert}=0.005$ and 0.0105, respectively. The perturbation frequencies are represented by white dashed lines. In all cases, the perturbation amplitude is set to $\xi^*_{\text{pert}}=0.01$. The power amplitude $A$ at each $x^*$ location is normalized by its local maximum $A_{\text{max}}(x^*)$, such that $A^* = A/A_{\text{max}}$.}
	\label{fig:Effect_pert_freq}
\end{figure}

\begin{figure}[tbp]
  \centering
  \includegraphics[width=1.0\textwidth, trim = {1.7cm 3.1cm 2.2cm 4.5cm},clip]{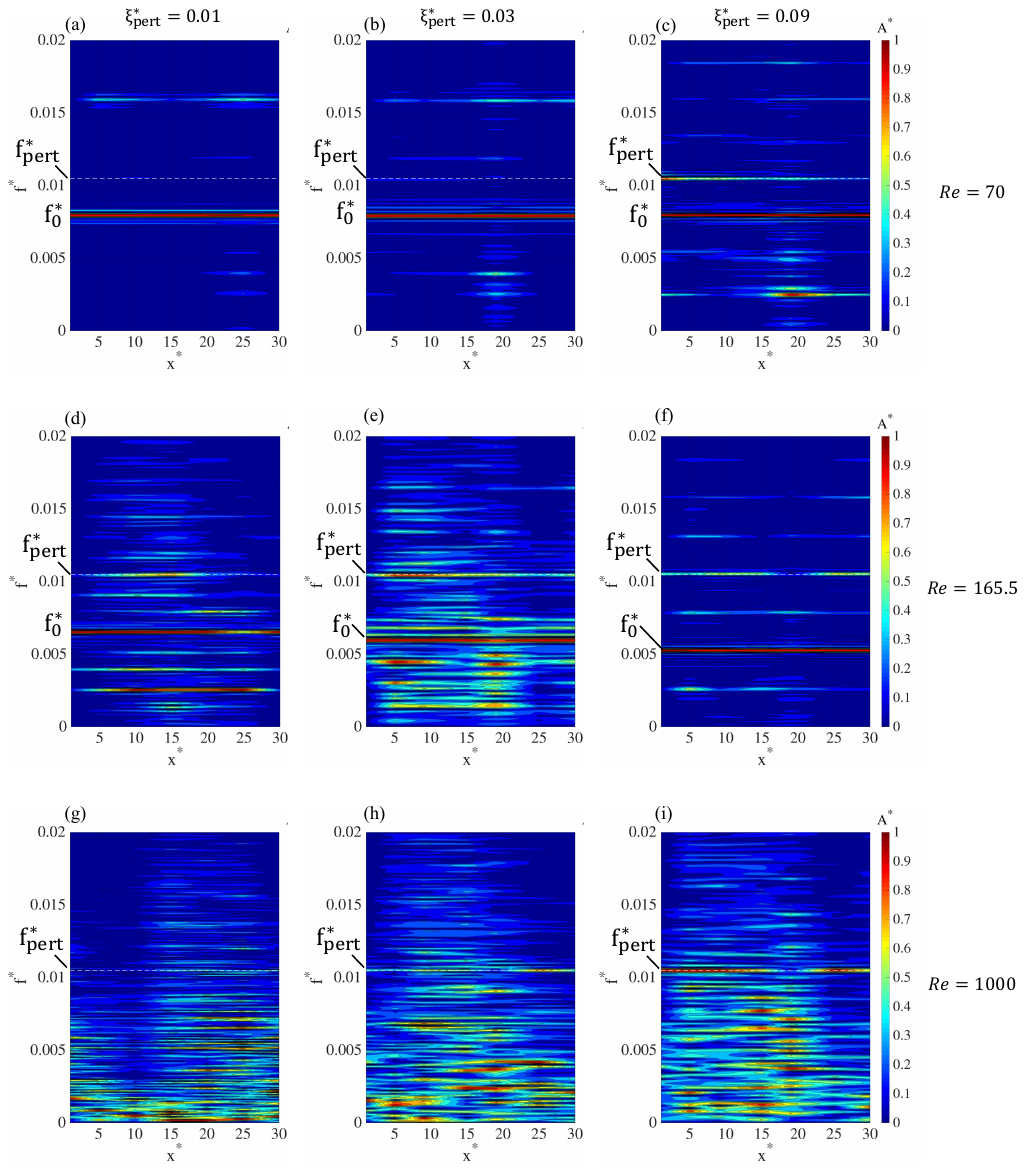}
	\caption{Spectrograms for different instability regimes. The first, second, and third rows correspond to $\text{Re}=70$ (absolute), 165.5 (weak absolute), and 1000 (convective), respectively. Different columns are for different inlet perturbation amplitudes: the first, second and third columns are for $\xi^*_{pert}=0.01$, 0.03, and 0.09, respectively. The perturbation frequency is set to $f^*_{\text{pert}}=0.0105$ for all cases and is represented by white dashed lines. The power amplitude $A$ at each $x^*$ location is normalized by its local maximum $A_{\text{max}}(x^*)$, such that $A^* = A/A_{\text{max}}$.}
\label{fig:Effect_pert_amp}
\end{figure}

To better understand the distinctive characteristics of each regime, we introduce perturbations to the liquid velocity at the inlet, see Eq.~\eqref{eq:pert_sine}. The spectrograms for different inlet perturbation frequencies and amplitudes are shown in Figs.~\ref{fig:Effect_pert_freq} and \ref{fig:Effect_pert_amp}, respectively. To better visualize the locally dominant frequency, the spectrum amplitude $A(x^*,f^*)$ is normalized by the local maximum, namely $A^* = A/A_{\max}$, where $A_{max}(x^*)=\max[A(x^*,f^*)]$. In Fig.~\ref{fig:Effect_pert_freq}, the perturbation amplitude is fixed at a small value, \ie, $\xi^*_{pert}=0.01$, while the perturbation frequencies are varied, including multi-mode perturbations in the first column with \tcr{10} different frequencies randomly distributed between $f^*_{pert} = 0.0015$ and 0.015 \tcr{(see Figs.~\ref{fig:Effect_pert_freq}(a,d,g))}, and single-mode perturbations with $f^*_{pert} = 0.005$ \tcr{(see Figs.~\ref{fig:Effect_pert_freq}(b,e,h))} and 0.0105 \tcr{(see Figs.~\ref{fig:Effect_pert_freq}(c,f,i))} in the second and third columns, respectively. These two frequencies are chosen to be smaller and larger than the absolute frequency for $\text{Re} = 70$. In contrast, for the results shown in Fig.~\ref{fig:Effect_pert_amp}, the perturbation frequency is fixed at $f^*_{pert} = 0.0105$, while the perturbation amplitude is varied, \ie, $\xi^*_{pert}=0.01$ \tcr{(see Figs.~\ref{fig:Effect_pert_amp}(a,d,g))}, 0.03  \tcr{(see Figs.~\ref{fig:Effect_pert_amp}(b,e,h))}, and 0.09 \tcr{(see Figs.~\ref{fig:Effect_pert_amp}(c,f,i))}. 

\subsubsection{Absolute instability regime}
When the shear instability at the interface is absolutely unstable, the system behaves like an oscillator. \tcr{A dominant frequency ($f^*_0=0.008$) appears as a red horizontal line in the spectrogram, as shown in Fig.~\ref{fig:Effect_pert_freq}(a), indicating that $f^*_0$ remains constant in space. Furthermore, it is evident that $f^*_0$ differs from all the frequencies of the perturbations induced at the inlet, which are represented by the white dashed lines. This suggests that the frequency of the dominant absolute mode is independent of the inlet perturbation frequency.} The results for single-mode perturbations shown in Figs.~\ref{fig:Effect_pert_freq}(b) and (c) are similar, in which the dominant frequency is the same as observed in Fig.~\ref{fig:Effect_pert_freq}(a) and is different from the perturbation frequencies.

As can be seen from the first row of Fig.~\ref{fig:Effect_pert_amp}, when the perturbation amplitude increases from $\xi^*_{pert} = 0.01$ to 0.03 and 0.09, the dominant frequency remains the same, indicating that the dominant frequency is unaffected by the inlet perturbation amplitude. Even with the highest perturbation amplitude $\xi^*_{pert} = 0.09$, for which the perturbation mode can be clearly seen near the inlet, the absolute mode remains dominant. As the perturbation amplitude is large, like $\xi^*_{pert} = 0.09$, it induces modes of lower and higher frequencies due to the nonlinear effect. The induced mode of low frequencies can grow spatially and become quite significant downstream, see $x^*\approx 20$.
 
\subsubsection{Convective instability regime}
The case $\text{Re}=1000$ is in the convective regime; there is no frequency that dominates the entire domain. Since multiple spatial modes grow, the spectrograms are very noisy, see Figs.~\ref{fig:Effect_pert_freq}(g-i). Furthermore, since convectively unstable systems are highly sensitive to inlet perturbations, the spectrograms for $\text{Re}=1000$ with multi-mode and single-mode perturbations are different. In Fig.~\ref{fig:Effect_pert_freq}(g), all perturbation modes induced at the inlet grow spatially. The modes induced by the perturbation modes also grow spatially due to the nonlinear effect, making the spectrogram even noisier. \tcr{Similar noisy spectrograms are also shown in previous study to indicate the instability is convective, see \eg, Bozonnet \etal \cite{Bozonnet_2022a}.}

Upon perturbing the inlet velocity with a low frequency of $f^*_{pert} = 0.005$, this mode emerges near the inlet. With the continued propagation of the wave, this mode experiences growth and becomes dominant for $x^*>15$, see Fig.~\ref{fig:Effect_pert_freq}(h). In contrast, for a perturbation with a high frequency, $f^*_{pert} = 0.0105$, though the mode can be recognized, its amplitude is pretty low, see Fig.~\ref{fig:Effect_pert_freq}(i). When the perturbation amplitude for $f^*_{pert} = 0.0105$ increases to $\xi^*_{pert}=0.09$, it dominates the modes of lower frequencies. The results seem to indicate the modes within the low-frequency range exhibit higher spatial growth rates, though the detailed spatial growth rates are hard to measure from the simulation results. Therefore, perturbation at a lower frequency is more effective in destabilizing the flow.

It should be noted that even for cases in which no explicit perturbations are introduced, the interference of the splitter plate with the flow also induces small perturbations at the inlet. As these perturbations grow spatially as well, they will also contribute to the noise observed in the spectrogram.

\subsubsection{Weak absolute instability regime}
Finally, we will discuss the case $\text{Re}=165.5$, which is in the weak absolute regime. This regime is a transition from the absolute to convective regimes and thus exhibits features of both, see Figs.~\ref{fig:Effect_pert_freq}(d-f). On one hand, the spectrograms exhibit the absolute frequency at $f^*_0=0.0067$, which is different from the perturbation frequencies, and the absolute mode generally dominates the whole $x^*$ range. On the other hand, the perturbation modes other than the absolute mode grow spatially as well, making the spectrograms somewhat noisy, similar to the convective instability case ($\text{Re}=1000$). In this transition regime, the temporal growth rate of the absolute mode is low, as revealed in the LSA, see Fig.~\ref{fig:sp_properties_Re}(a). Therefore, even though it dominates, its amplitude may be comparable to the convective modes. For some spatial regions, like $x^*=25$, the amplitude of the convective mode even exceeds that of the absolute mode.

 Another interesting feature of the weakly absolute regime can be observed in Figs.~\ref{fig:Effect_pert_amp}(d-f), i.e., the dominant frequency decreases as the perturbation amplitude increases. The absolute frequency $f^*_0$ decreases from 0.0067 to about 0.0052 when $\xi^*_{pert}$ increases from 0.01 to 0.09. \tcr{Such a variation of $f^*_0$ with perturbation amplitude was not observed in the absolute instability regime ($\text{Re}=70$).} Since the perturbation frequency for the cases shown in Fig.~\ref{fig:Effect_pert_amp} is $f^*_{pert}=0.0105$, when its amplitude is high, it tends to induce a secondary mode at its half frequency $f^*_{pert}\approx 0.005$, which is quite close to the absolute frequency without perturbation. \tcr{The shift in the dominant frequency $f^*_0$ may be due to the merging of the original absolute mode with the induced mode.} In general, due to the relatively low temporal growth rate and amplitude, the absolute mode in the weak absolute regime is more sensitive to the nonlinear effect and perturbation amplitude.

\subsubsection{Variation of absolute frequency with Re}
\begin{figure}[tbp]
  \centering
  \includegraphics[width=1\textwidth, trim = {4.7cm 11.5cm 4cm 11cm},clip]{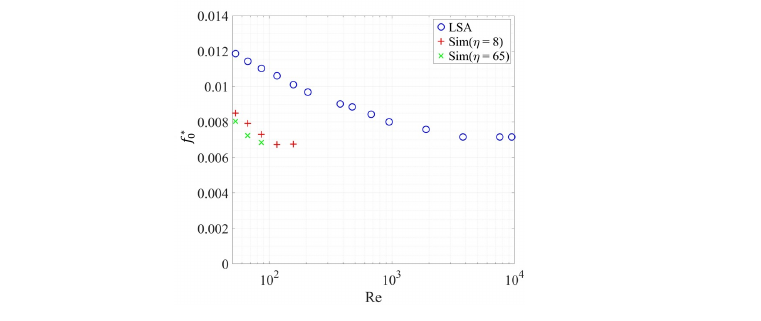}
\caption{Variation of dominant frequency with  $\text{Re}$ for $\eta=8$ and 65. The LSA predictions are shown for comparison.}
\label{fig:freq_Re_eta8}
\end{figure}

The variation of the dominant frequency \tcr{$f^*_0$} with $\text{Re}$ in the absolute and weak absolute regimes is shown in Fig.~\ref{fig:freq_Re_eta8}. For the case in the weak absolute regime, since the absolute frequency can vary with the perturbation amplitude, here we show the value for no perturbation at the inlet. The dominant frequencies predicted by LSA (section \ref{sec:LSA}) are also shown for comparison.

Both simulations and LSA results indicate that the frequency decreases with $\text{Re}$. The decrease of $f$ slows down and reaches a plateau in the weak absolute regime before the instability transitions to the convective regime. Furthermore, both results show that the interfacial instability transitions from absolute to convective regimes when $\text{Re}>\text{Re}_{cr}$. Near $\text{Re}_{cr}$, the dominant frequency is about $f^*_0\approx 7\times 10^{-3}$, and both simulations and LSA yield similar predictions.

The discrepancy between simulations and LSA lies in the value of $\text{Re}_{cr}$. For the simulations, $\text{Re}_{cr} \approx 165.6$, which is much lower than the prediction of LSA, which is about $\text{Re}_{cr} = 10,000$. As a result, for a given $\text{Re} < \text{Re}_{cr}$, the LSA frequency is lower than the simulation result. Similar underestimates of frequency by LSA have also been reported in previous works \cite{Matas_2015a}. The discrepancy between LSA and simulations is unlikely due to the confinement effect, which will be discussed in the next section. The discrepancy is also not due to the velocity profile of the base flow used in LSA, since $\text{Re}_{cr}$ varies little when $U_0$ increases from 0 to 0.1, see Appendix~\ref{sec:effect_infc_vel}. The difference may be related to the fact that LSA does not consider the spatial variations of the interfacial velocity and the base flow velocity profile. In the full simulations, the interfacial velocity increases and the velocity deficit reduces along the streamwise direction. Furthermore, the interfacial waves grow in amplitude as they propagate downstream, so the nonlinear wave dynamics and interaction between the wave and the gas stream downstream will unavoidably influence the interfacial instability in the near field. Nevertheless, a more thorough investigation is required for future work to fully confirm the reason behind the discrepancy.

\subsection{Effect of gas viscosity on far-field wave dynamics}
\label{sec:farfield}
\tcr{The results of simulation series A (confined configuration) are presented to illustrate the effect of gas viscosity on far-field wave dynamics.}
\subsubsection{Wave propagation}
\begin{figure}[tbp]
  \centering
  \includegraphics[width=1.\textwidth, trim = {1.5cm 7.5cm 2.5cm 6.8cm},clip]{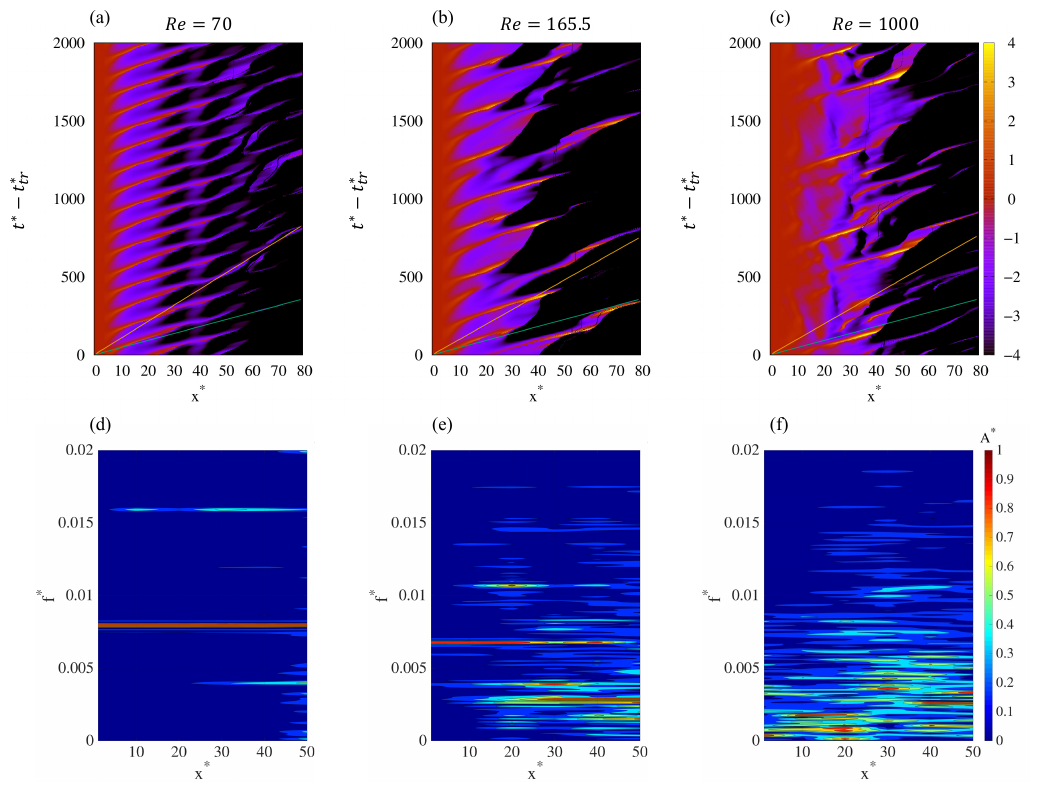}
\caption{(a)-(c) Spatio-temporal diagrams of the interface height and (d)-(f) spectrograms for cases with different $\text{Re}$ in Series A. The yellow and green lines indicate the celerity $U^*_c$ predicted by LSA and the Dimotakis speed $U^*_D$, respectively. }
\label{fig:far_field_dynamics}
\end{figure}

The propagation of interfacial waves induced by the interfacial instability is visualized by the spatial-temporal diagrams of interfacial height, see Fig.~\ref{fig:far_field_dynamics}. For the case $\text{Re}=70$ in the absolute regime, the trajectories of individual waves are equally spaced in $t^*$, indicating they follow the absolute frequency. The slope of the wave trajectory represents the inverse of the wave propagation speed, which is similar for all waves shown here. It is further observed that the wave propagation speed varies in $x^*$. The wave speed is lower near the inlet ($x^*\lesssim 10$), where the wave amplitude remains low. The wave speed is found to be quite similar to the celerity $U_c^* = \omega_r^*/\alpha_r^*$ predicted by the LSA (see the yellow line). As the wave propagates further downstream, the wave speed increases and eventually reaches a constant for $x^* \gtrsim 15$, which is similar to the Dimotakis speed $U_D^*$, indicated by the green line. The good agreement between the wave speed and the Dimotakis speed has been reported in previous studies \cite{Ling_2017a, Jiang_2021a, Bozonnet_2022a}, however, less attention has been paid to the fact that the wave speed is actually lower near the inlet, though the wave speed variation can also be recognized from the previous results \cite{Jiang_2021a, Bozonnet_2022a}.

It is worth noting that the Dimotakis speed is an inviscid feature, while the celerity $U_c^*$ is predicted by the viscous stability analysis and thus involves the viscous effect. The present results indicate that the viscous effect is important in the selection of the most unstable mode and the propagation of the interfacial wave in the near field, and the propagation of the wave further downstream is controlled mainly by the inviscid mechanism.

Similar observations can be made for the case $\text{Re}=165.5$, which is in the weak absolute regime. The wave speed agrees with $U_c^*$ in the near field and matches with $U_D^*$ in the far field. Since the case $\text{Re}=1000$ is in the convective instability regime, it is difficult to identify individual waves in the near field. Nevertheless, it is observed that the wave propagation speed agrees well with $U_D^*$ in the far field where the wave trajectories are better seen.

\subsubsection{Nonlinear wave-wave interaction}
As the amplitude of the interfacial waves grows spatially, the nonlinear effect intensifies. Figure~\ref{fig:Flow_behaviour_confined} shows how interfacial waves, upon interacting with the gas stream, generate liquid sheets that extend from wave crests and eventually break into small droplets. The detailed droplet formation mechanisms were revealed in previous 3D simulations \cite{Ling_2017a, Ling_2019b, Agbaglah_2021a, Ling_2023a}. Upon wave breakup, the interfacial height significantly drops to a lower value, corresponding to the remaining unbroken liquid layer near the bottom, that is why the waves seem to ``disappear" in the spatial-temporal diagrams (see Figs.~\ref{fig:far_field_dynamics}(a)-(c)).

Before wave breakup, a decrease in propagation speed is observed, which is due to the flow separation downstream of the wave. As a result, the subsequent wave may catch up and merge with the former one. This wave-wave interaction is most profound for the case $\text{Re}=165.5$, see Fig.~\ref{fig:far_field_dynamics}(b).  A significant outcome of this wave interaction is the reduction of the dominant frequency in the spectrum from $f^*_0=0.0067$ in the near field to approximately half ($f^*_0=0.0038$) downstream (Fig.~\ref{fig:far_field_dynamics}(b)). This finding implies that although the wave formation frequency near the inlet is governed by the absolute instability, the frequency observed downstream can be significantly lower. The variation of dominant frequency spatially due to wave-wave interactions highlights the importance to show the whole spectrogram.

In the $\text{Re}=70$ case, waves merging is not observed, likely due to the generally lower wave amplitude and minor wave speed variations. However, nonlinear effect is still evident in the spectrogram, which reveals secondary modes at frequencies that are integer multiples and half of the absolute mode's frequency (Fig.~\ref{fig:far_field_dynamics}(d)). For $\text{Re}=1000$, where the flow exhibits convective instability, the absence of a singular dominating mode makes it challenging to discern individual waves and their interactions.

\subsection{Effect of confinement}
\label{sec:confinement}
The results presented above are for the first configuration of the two-phase mixing layer (Fig.~\ref{fig:Computational_domains}(a)) with a small $\eta=8$, therefore, the effect of confinement may influence the interfacial instability and interfacial wave development. In this section, we will present the results for \tcr{the unconfined configuration, namely series B with a large $\eta=65$, see Fig.~\ref{fig:Computational_domains}(b). The comparison between the results for Series A and B  will thus serve to characterize the effect of confinement.}

\subsubsection{Spatial growth of wave amplitude}
\begin{figure}[tbp]
  \centering
  \includegraphics[width=0.99\textwidth, trim = {1.3cm 9.5cm 1.5cm 6.8cm},clip] {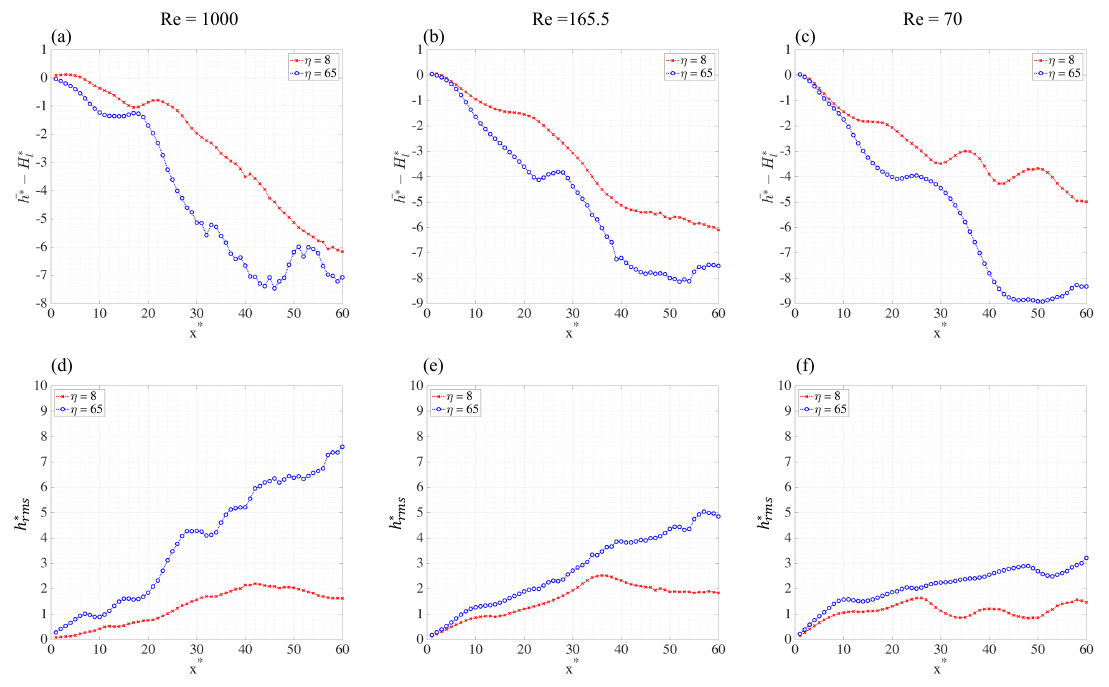}
\caption{Spatial development of (a)-(c) mean interfacial height $\bar{h}^*$ and (d)-(f) wave amplitude $h^*_{rms}$  for different $\text{Re}$ and $\eta$.}
\label{fig:Spatial_growth}
\end{figure}

The most direct impact of confinement is on the spatial development of the liquid stream thickness, represented by the time-average interfacial height, 
\begin{equation}
   \bar{h^*}(x) = \frac{1}{t_{ave}} \int_{t_{tr}}^{t_{tr}+t_{ave}}  h^*(x,t) \, dt \,,
\end{equation}
and the the interfacial wave amplitude, represented by the root-mean-square (rms) of interfacial height fluctuations
\begin{equation}
	{h}^*_{rms}(x) = \sqrt{\frac{1}{t_{ave}} \int_{t_{tr}}^{t_{tr}+t_{ave}} \left( h^*(x,t) - \bar{h}^* \right)^2\, dt }\,.
	\label{eq:wave_amp_def}
\end{equation}
Here, $t_{ave}$ and $t_{tr}$ represent the time duration for averaging and the transition time for the two-phase mixing layer to reach a statically steady state. The spatial variations of $\bar{h}^*$ and ${h}^*_{rms}$ for different $\eta$ and $\text{Re}$ are shown in Fig.~\ref{fig:Spatial_growth}. As the liquid stream is accelerated by the gas counterpart and the liquid velocity increases along $x$, the mean liquid stream thickness, $\bar{h}^*$, decreases over $x$ due to mass conservation. For a given $\text{Re}$, $\bar{h}^*$ for the unconfined case decreases faster and reaches a lower value. The spatial development of $\bar{h}^*$ also varies with $\text{Re}$ and the trends are similar for both confined and unconfined cases near the inlet. For low $\text{Re}=70$, $\bar{h}^*$ decreases more rapidly in $x$, compared to the case with high $\text{Re}=1000$. The increase of gas viscosity enhances the transverse momentum transport between the gas and liquid streams and the acceleration of the liquid, which in turn leads to the more rapid decrease of liquid stream thickness. 

The spatial development of the interfacial wave amplitude $h^*_{rms}$ shows similar variation trends when $\text{Re}$ varies. It is seen that $h^*_{rms}$ for $\text{Re}=70$ increases faster in $x^*$ near the inlet, indicating that the enhanced momentum transport also contributes to a higher spatial growth rate of the interfacial wave. Nevertheless, an opposite trend is observed further downstream, where $h^*_{rms}$ grows faster for $\text{Re}=1000$ and reaches higher values. For $\text{Re}=70$, the larger gas velocity contributes to a higher energy dissipation, the interaction between the gas stream and interfacial wave downstream becomes weaker, see Fig.~\ref{fig:Flow_behaviour_confined}. In contrast, the interfacial wave for $\text{Re}=1000$ continues to grow and has a strong interaction with the gas stream downstream, leading to wave rolling and breakup. 

It is important to note that for the confined case, the effect of confinement is substantial when ${h}^*_{rms}$ is still lower than the liquid layer thickness, $H^*_l=8$. As expected, the wave amplitude grows more rapidly with $x^*$ and reaches higher values for the case $\eta=65$.

\subsubsection{Convective-absolute instability transition}
\begin{figure}[tbp]
  \centering
  \includegraphics[width=1.1\textwidth, trim = {1.55cm 7.5cm 2cm 6.8cm},clip]{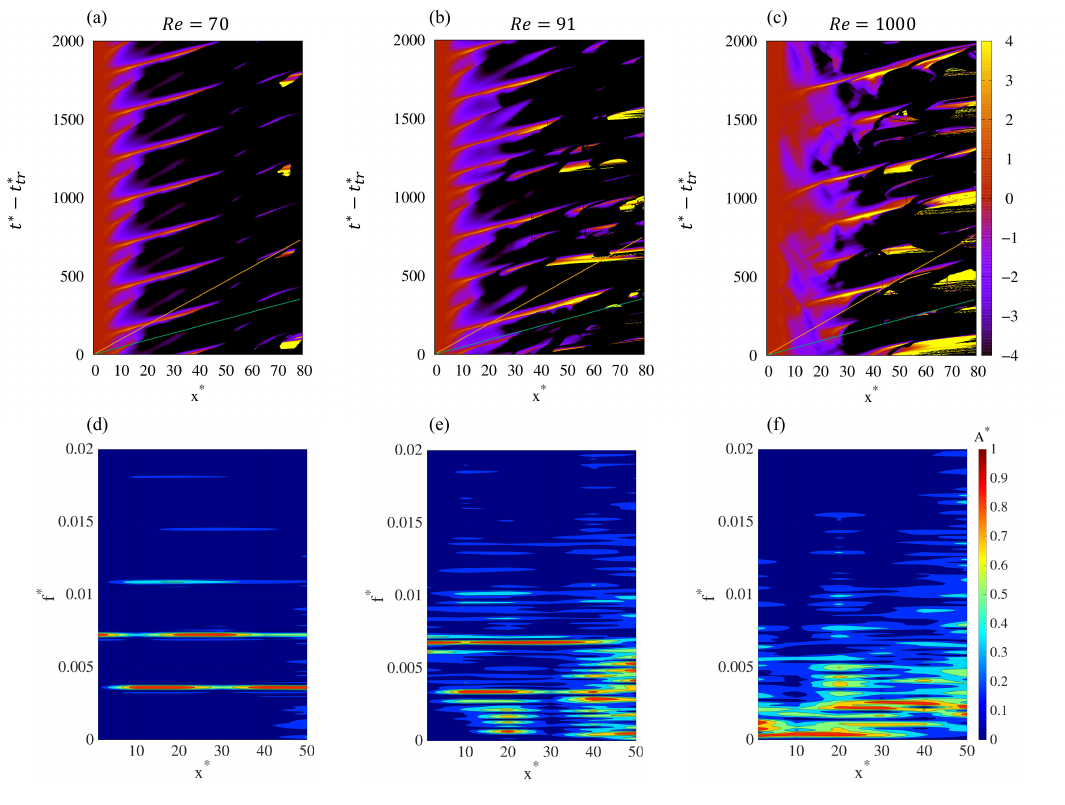}
\caption{(a)-(c) Spatio-temporal diagrams of the interface height and (d)-(f) spectrograms for cases with different $\text{Re}$ in Series B. The yellow and green lines indicate the celerity $U^*_c$ predicted by LSA and the Dimotakis speed $U^*_D$, respectively. }
\label{fig:far_field_dynamics_unconfined}
\end{figure}

\tcr{The A/C transition induced by $\text{Re}$ observed for the confined configuration is also present in the unconfined counterpart $(\eta = 65)$. The different regimes, including the convective, weak absolute, and absolute instability regimes, are clearly identified from the spatio-temporal evolutions of the interfacial height shown in  Fig.\ref{fig:far_field_dynamics_unconfined}.} 
For the case $\text{Re} = 70$, individual wave trajectories can be clearly identified in the near field, indicating that the instability is in the absolute regime. Yet, different from the confined case with the same $\text{Re} = 70$, more profound wave-wave interaction is observed, which is due to faster spatial growth and the resulting elevated nonlinear effect. The outcome of this nonlinearity is the shift of the dominant frequency, from $(f^*_0 = 0.0072)$ to almost half of the dominant frequency, $(f^*_0 = 0.0036)$. The case $\text{Re} = 91$ is likely in the weak absolute regime, since the spectrogram shows both the absolute mode and the noise due to convective modes. Nonlinear wave-wave interaction is also observed for this case. For $10 \le x^* \le 30$, the secondary mode (with frequency about half of the absolute mode) induced by the nonlinear effect carries comparable energy to the absolute mode. For the case $\text{Re} = 1000$, wave trajectories cannot be identified in the near field and the spectrogram is very noisy, indicating that this case is in the convective regime.

The spatial variation of the wave propagation speed for the unconfined cases is also similar to that observed for the confined cases. For both $\text{Re} = 70$ and 91, the wave speed in the near field follows the celerity $U^*_c$ predicted by LSA, and the wave speed increases to $U^*_D$ further downstream. When $\text{Re}$ increased to $\text{Re} = 1000$, similar to the confined domain, waves at the inlet are difficult to identify, but the wave speed agrees well with $U^*_D$ downstream. Therefore, the viscous effect plays a significant role in the formation and propagation of interfacial waves in the near field for both confined and unconfined cases.

\begin{figure}[tbp]
 \centering
 \includegraphics[width=0.99\textwidth, trim = {1.1cm 9.5cm 1.5cm 6cm},clip] {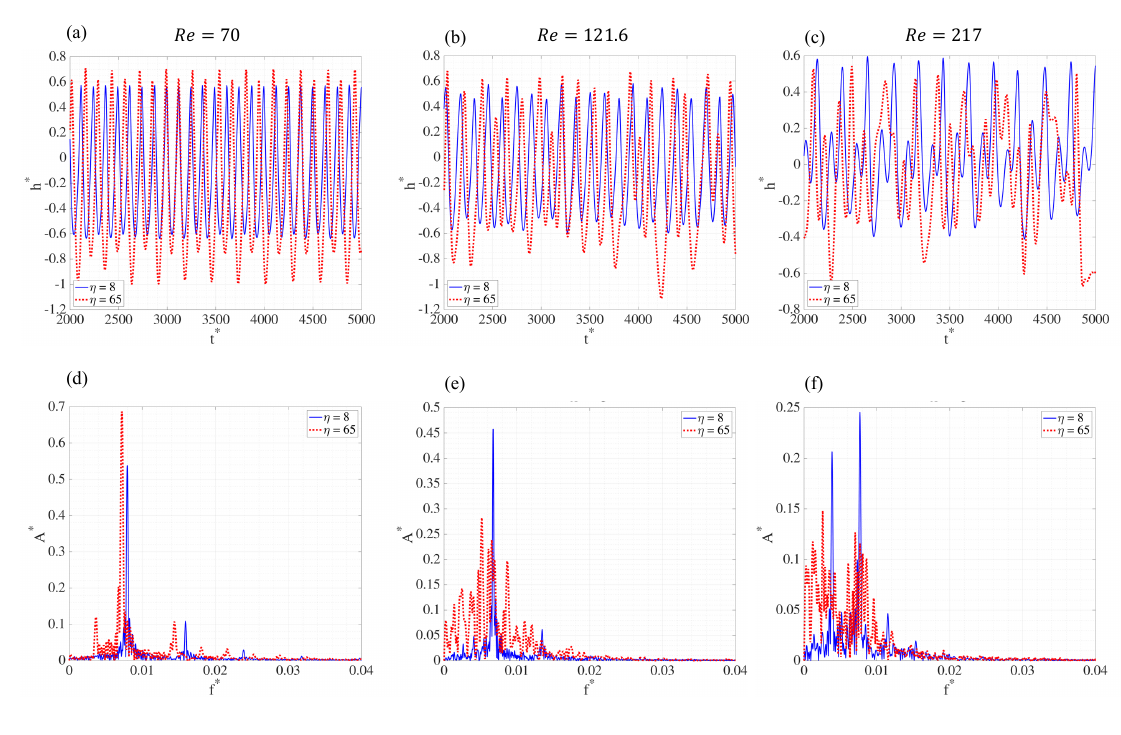}
	\caption{(a)-(c) Temporal evolutions and (d)-(f) frequency spectra of the interfacial height at $x^* = 3$ for different $\text{Re}$ and $\eta$. }
\label{fig:Confinement_effect} 
\end{figure}

To better illustrate the effect of $\eta$ on the interfacial instability, the temporal evolutions and frequency spectra of the interfacial height at $x^* = 3$ for different $\text{Re}$ are shown in Fig.~\ref{fig:Confinement_effect}. For each $\text{Re}$, the results for different $\eta$ are plotted. For the case $\text{Re}=70$, both the confined and unconfined cases are in the absolute regime, and the oscillations of the interfacial height are very similar. The oscillation amplitude for $\eta=65$ is slightly larger, due to the faster spatial growth of the wave amplitude. The dominant frequencies in the spectra are also similar, with the frequency for $\eta=65$ being slightly lower. This indicates that the confinement effect on the absolute frequency is minor, for the range of $\eta$ considered. The spectrum for $\eta=65$ shows a stronger nonlinear effect, and secondary modes at half and multiple times of the frequency of the absolute mode are stronger. 

For the case $\text{Re}=121.6$, while the confined case $\eta=8$ is still in the absolute regime, showing a clear dominant frequency, the unconfined case is in the weak absolute regime, for which the amplitudes of the convective modes are comparable to that of the absolute mode. When $\text{Re}$ continues to increase to 217, while the confined case is in the weak absolute regime, the unconfined case transitions to the convective regime, showing a noisy spectrum. The results indicate that the A/C transition occurs at a smaller $\text{Re}_{cr}$ for the unconfined configuration when the confinement effect is reduced. \tcr{It is measured that $\text{Re}_{cr}\approx 90$ for the unconfined configuration and $r=0.05$, which is about 45\% lower than the confined counterpart.}

\subsubsection{Variation of absolute frequency over Re}
\tcr{
The absolute frequencies $f^*_0$ for both the absolute and weak absolute regimes in the confined and unconfined configurations are shown in Fig.~\ref{fig:freq_Re_eta8}, alongside the LSA predictions for the unconfined configuration. Interestingly, the discrepancy between the simulation results for the unconfined configuration and the LSA predictions is greater than that for the confined configuration. This indicates that the observed differences between the simulations and LSA predictions are not related to the confinement effect.
}

Former LSA studies \cite{Matas_2015a, Bozonnet_2022a} show that absolute instability can be induced by the pinching between the shear and confinement branches. Though the present cases do not belong to that category of absolute instability, the simulation results indicate that the confinement effect can influence the critical $\text{Re}$ for A/C transition.

Finally, it is observed that $f^*_0$ for both confined and unconfined configurations decreases with $\text{Re}$. The values of $f^*_0$ for the unconfined configuration are slightly lower (about 5\%) than their confined counterparts at the same $\text{Re}$. In general, for the confined configuration with $\eta=8$ considered, the confinement effect is mainly on changing the critical $\text{Re}$. For cases in the absolute regime, the effect of confinement on the dominant frequency is minor.

\subsection{Effect of density ratio}
\label{sec:densityratio}
\tcr{In simulation series A}, the gas-to-liquid density ratio $r$ is fixed at 0.05, which is relatively large, compared to typical gas-liquid systems. In the simulation series C \tcr{(confined configuration)}, see Table \ref{tab:cases}, the density ratio is reduced to $r = 0.0125$, which is one fourth of the value for series A. The value of $r$ is decreased by reducing the gas density $\rho_g$, see Table \ref{tab:phy_para1}, while other parameters, like $\eta$, are kept as the same as simulation series A. Similar to other simulation series, the gas viscosity is varied to investigate the effect gas viscosity. 

\begin{figure}[tbp]
  \centering
  \includegraphics[width=0.99\textwidth, trim = {1cm 7cm 1cm 6.5cm},clip] {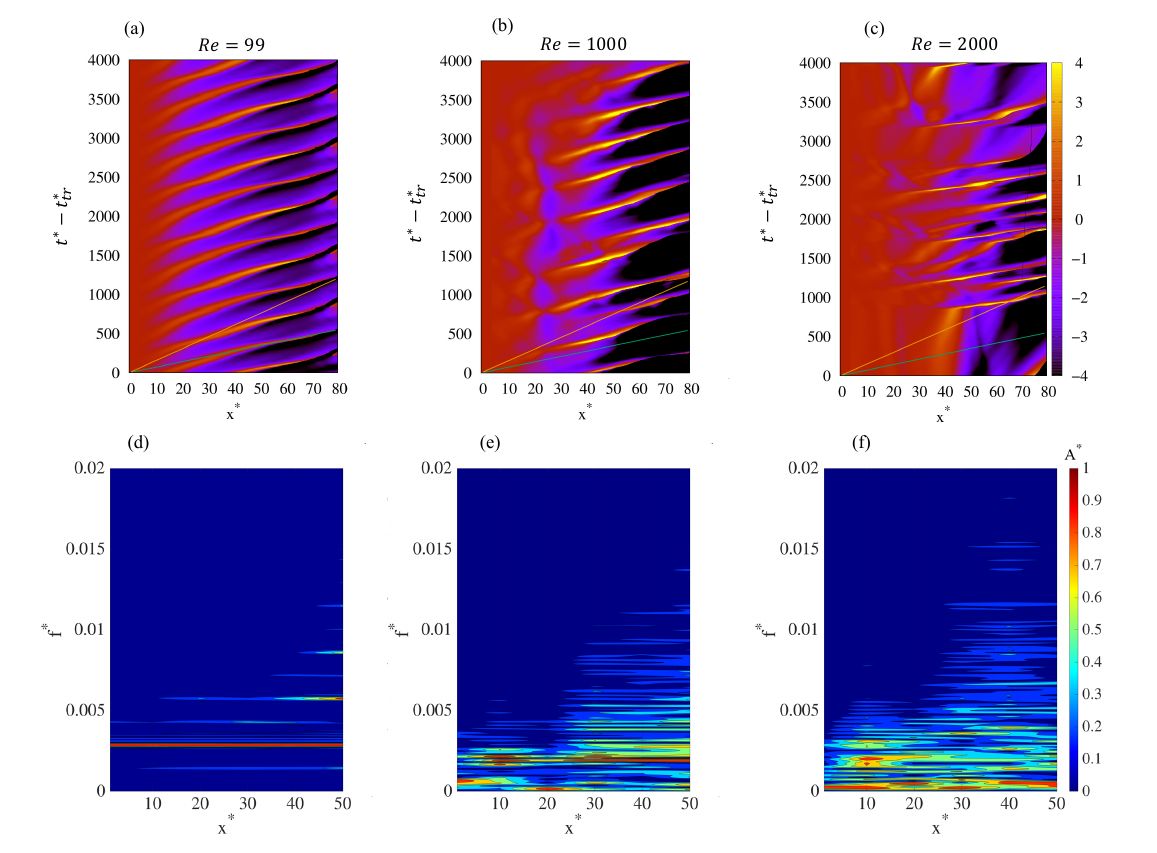}
\caption{(a)-(c) Spatio-temporal diagrams of the interface height and (d)-(f) spectrograms for cases with different $\text{Re}$ in Series C. The yellow and green lines indicate the celerity $U^*_c$ predicted by LSA and the Dimotakis speed $U^*_D$, respectively. }
\label{fig:Effect_density_ratio}
\end{figure}

Figure~\ref{fig:Effect_density_ratio} show the spatio-temporal diagrams for the interfacial height and the corresponding spectrograms for different $\text{Re}$ and $r=0.0125$. Three distinct instability regimes can be clearly identified, and the key features for each regime are quite similar to those for $r=0.05$. The interfacial instability for $\text{Re} = 99$ is absolute, and the dominant frequency of the absolute mode is clearly seen in both the spatial-temporal diagram and the spectrogram. Individual waves can be well identified in Fig.~\ref{fig:Effect_density_ratio}(a). Similar to the results for $r=0.05$ shown in Fig.~\ref{fig:far_field_dynamics}(a), the wave speed is in good agreement with the $U_c^*$ (yellow line) in the near field and then increases to the Dimotakis speed, $U_D^*$ (green line) downstream. The absolute mode, with a frequency $f^*_0 = 0.003$ dominates the whole domain, see Fig.~\ref{fig:Effect_density_ratio}(d), though secondary modes are also present induced by nonlinear effect. 

In contrast, for the case with large $\text{Re} = 2000$, the instability is convective, as a result, the spectrogram is noisy. It is again difficult to identify individual waves in the spatio-temporal diagram, see Fig.~\ref{fig:Effect_density_ratio}(c), but the wave speed in the far field agrees well with $U_D^*$. For the intermediate $\text{Re} = 1000$, the spectrogram shows a dominant mode accompanied by other convective modes, which indicates the case is in the weak absolute regime. The main effect of $r$ on the interfacial instability lies at the critical $\text{Re}$ for the C/A transition. The critical value for $r = 0.0125$ is about 1000, which is higher than the value 217 for $r = 0.05$.

\begin{figure}[htbp]
  \centering
  \includegraphics[width=0.99\textwidth, trim = {4cm 15.5cm 4cm 7cm},clip]{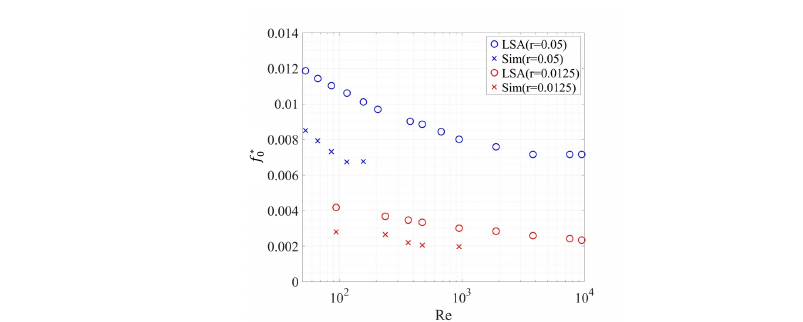}
\caption{Variation of dominant frequency $f^*_0$ with  $\text{Re}$ for gas-to-liquid density ratios $r = 0.05$ and $0.0125$. The simulation results are compared with the LSA predictions.}
\label{fig:freq_Re_r0p0125}
\end{figure}

The simulation results for $f^*_0$ as a function for $\text{Re}$ for $r=0.0125$ and 0.05 are shown in Fig.~\ref{fig:freq_Re_r0p0125}. For both $r=0.05$ and 0.0125, $f^*_0$ decreases with $\text{Re}$ and reaches a plateau when $\text{Re}$ reaches the critical value A/C transition. The values of $f^*_0$ are generally reduced when $r$ decreases from 0.05 to 0.0125. Beyond the decrease in $r$, the cases in Series C also exhibit a lower dynamic pressure ratio $M$, compared to the Series A, which also contributes to the lower $f^*_0$ observed. The LSA predictions for $r=0.0125$ are also shown for comparison, which are qualitatively similar to the simulation results, but the values of $f^*_0$ are again higher, similar to the cases with $r=0.05$. Furthermore, the value of $\text{Re}_{cr}$  predicted by the LSA varies little with $r$, while $\text{Re}_{cr}$ measured from the simulation results increases from about 160 to 1000, when $r$ decreases from 0.05 to 0.0125. 

\section{Conclusions}
\label{sec:conclusions}
The interfacial instability development in a two-phase mixing layer between two parallel planar gas and liquid streams have been studied through linear spatial-temporal viscous stability analysis and detailed numerical simulations. Parametric studies were performed using both the approaches to characterize the effect of gas viscosity on the interfacial instability. For the ranges of parameters considered, the absolute instability, if it occurs, belongs to the surface-tension type of absolute instability. The key finding of the present study is that when all other parameters are fixed, the instability can transition from absolute to convective (A/C) regimes when the gas viscosity decreases. As the gas viscosity decreases, the gas Reynolds number ($\text{Re}$) increases and the gas-to-liquid viscosity ratio ($m$) decreases. Linear stability analysis is conducted to identify the effects of $\text{Re}$ and $m$ separately. The results indicate that $\text{Re}$ plays a more important role in the transition. When $\text{Re}$ increases, the temporal growth rate decreases. When $\text{Re}$ reaches the critical value, the temporal growth rate becomes zero and the A/C transition occurs. The $\text{Re}$-induced A/C transition is also captured by the numerical simulations and can be clearly identified on in the spatial-temporal diagrams of the interfacial height and also the spectrograms. This conclusion that A/C transition occurs when $\text{Re}$ increases holds for both confined and unconfined configurations, which the ratio between the liquid stream height and the inlet gas boundary layer thickness $\eta$ varies from 8 to 65. The conclusion is also valid when the gas-to-liquid density ratio varies from 0.05 to 0.125. The effects of confinement and density ratio are shown to change the critical $\text{Re}$ for the A/C transition. \tcr{The critical $\text{Re}$ for $\eta=65$ (unconfined configuration) is approximately 45\% lower than that for $\eta=8$ (confined configuration). When the gas-to-liquid density ratio $r$ decreases from 0.05 to 0.0125, the critical $\text{Re}$ increases by about sixfold.} As the interfacial waves induced by the instability propagates and grow in amplitude, it is observed that wave speed increases from the celerity determined by the absolute mode to the Dimotakis speed. As the wave amplitude further grows to induce flow separation in the wake of the wave crest, the wave speed is reduced. For some cases, the subsequent wave can catch up and merge with the former one, and the wave merging results in frequency reduction. 

\begin{appendix}

\section{Effect of interfacial velocity in linear stability analysis}
\label{sec:effect_infc_vel}
\begin{figure}[tbp]
  \centering
   \includegraphics[width=0.99\textwidth, trim = {5cm 11cm 5cm 11cm},clip] {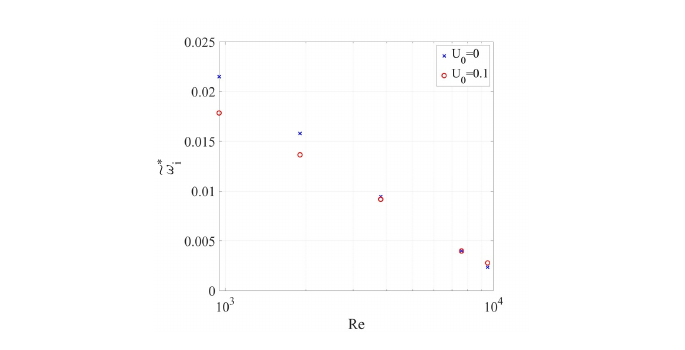}
\caption{LSA results for the temporal growth rate at the saddle point, $\tilde{\omega}_i^*$, as a function of $\text{Re}$ for different interfacial velocities $\text{U}_0$.}
\label{fig:Sensitivity_U0}
\end{figure}

The interfacial velocity $U_0$ is varied in the linear stability analysis (see Section \ref{sec:LSA}) to study its effect on the saddle-point feature. Figure~\ref{fig:Sensitivity_U0} shows the LSA results for the temporal growth rate $\tilde{\omega}^*_i$ for $U_0/U_l=0.1$ and 0. To keep $U_0$ constant when $\mu_g$ is varied, the value of $\delta_d$ needs to be adjusted accordingly, following Eq.~\eqref{eq:interface}. It is observed that the results for both $U_0$ values are similar, and $\omega^*_i$ decreases with $\text{Re}$. The values for $U_0/U_l=0.1$ are lower than those for the same $\text{Re}$. When $\omega^*_i$ reaches zero, the absolute-to-convective (A/C) transition occurs. The critical Reynolds number for both $U_0$ values are similar, implying that changes in the interfacial velocity do not affect the A/C transition and $\text{Re}_{cr}$ is not sensitive to the specific value of interfacial velocity.

\section{Effect of mesh resolution}
\label{sec:grid_refinement}
\begin{figure}[tbp]
  \centering
   \includegraphics[width=0.99\textwidth, trim = {3.1cm 9cm 5cm 6.6cm},clip] {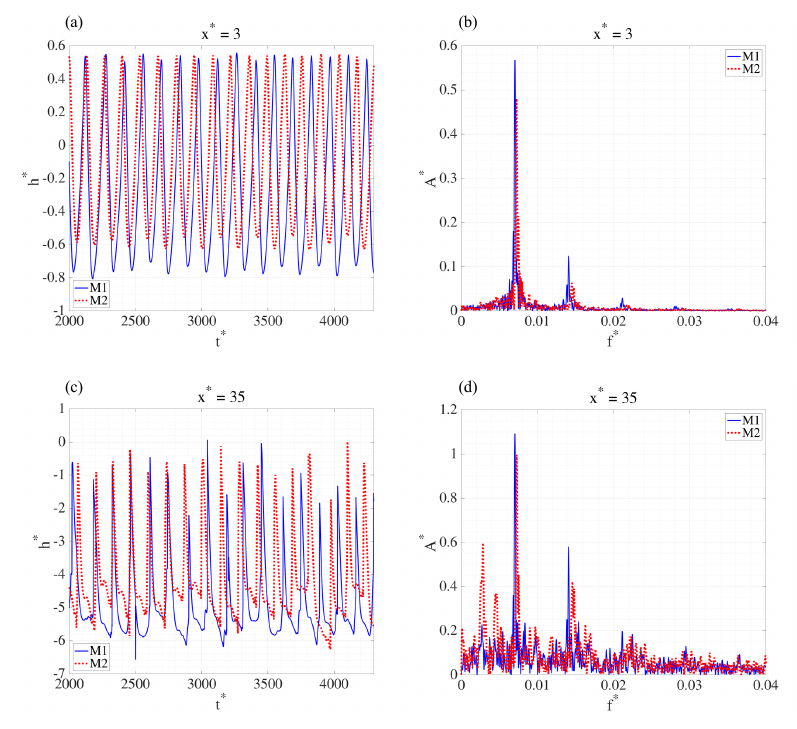}
  \caption{(a) Temporal evolutions and (b) frequency spectra of the interfacial height for different mesh resolutions and measurement locations. The meshes M1 and M2 represent $H/\Delta_x=128$ and 256, respectively. Representative locations are chosen, \ie, $x^* = 3$ and $35$. The case is for $\text{Re}=86$ and $\eta=8$ in series A.}   
\label{fig:Mesh_resolution_confined}
\end{figure}

\begin{figure}[tbp]
  \centering
   \includegraphics[width=0.99\textwidth, trim = {3.8cm 8.1cm 4.4cm 7.3cm},clip] {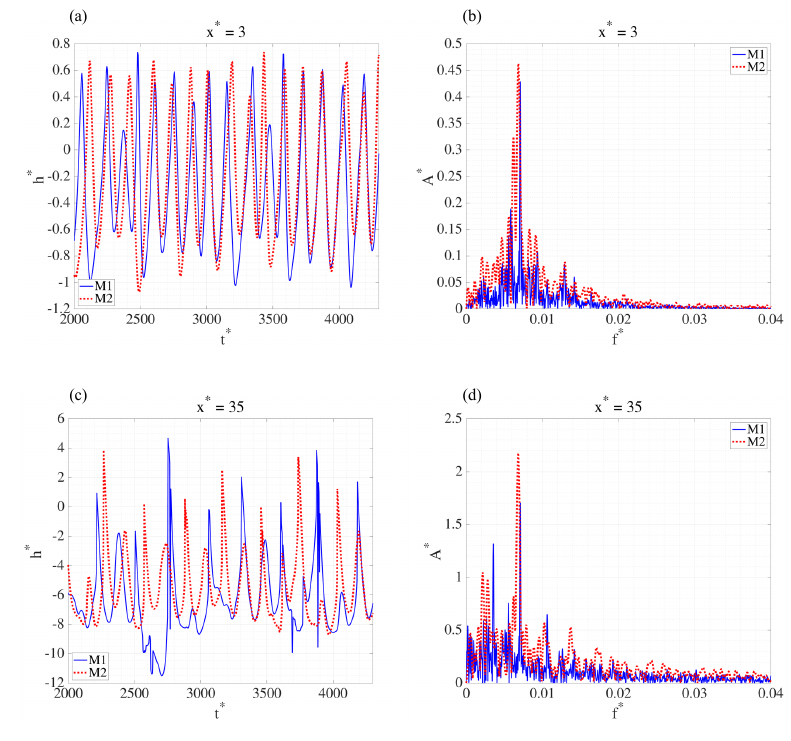}
  \caption{(a) Temporal evolutions  and (b) frequency spectra of interfacial height for different mesh resolutions and measurement locations. The meshes M1 and M2 represent $H/\Delta_x=128$ and 256, respectively. Representative locations are chosen, \ie, $x^* = 3$ and $35$. The case is for $\text{Re}=86$ and $\eta=65$ in series B.}   
\label{fig:Mesh_resolution_unconfined}
\end{figure}

The results for the grid convergence study are shown in Figs.~\ref{fig:Mesh_resolution_confined} and ~\ref{fig:Mesh_resolution_unconfined} for both the confined and unconfined configurations, i.e., $\eta=8$ and 65, respectively. The two mesh sizes used are referred to as M1 ($\Delta x=12.5\mu$m) and M2 ($\Delta x=6.25\mu$m). Since the convective cases are more complicated and don't show any dominant frequency, we have used an absolute case, $\text{Re} = 86$, for the grid refinement study. The time evolutions and frequency spectra for the interfacial height at two different locations are shown. It can be seen that the results for the interfacial height oscillation amplitude and frequency for the two meshes match well for both $\eta=8$ and 65, indicating that the M2 mesh is sufficient to yield grid-independent results.

\section{Effect of simulation time}
\label{sec:sim_time_study}
\begin{figure}[tbp]
  \centering
   \includegraphics[width=0.99\textwidth, trim = {3.1cm 12cm 5cm 10cm},clip] {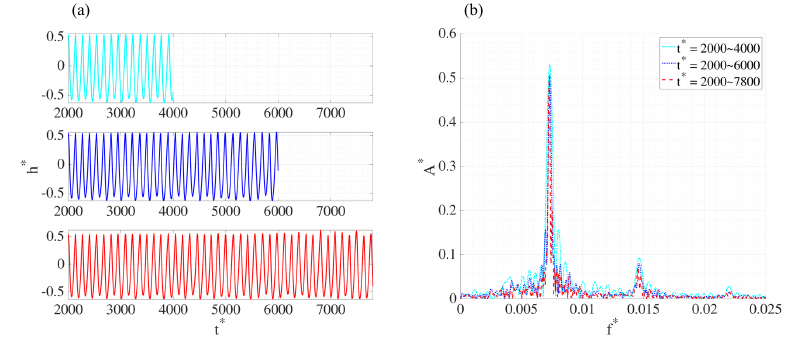}
  \caption{(a) Temporal evolutions  and (b) frequency spectra of interfacial height measured at $x^* = 3$ for different simulation  times. The case is for $\text{Re}=86$ and $\eta=65$ in series A.}   
\label{fig:Time_study_confined}
\end{figure}

\begin{figure}[tbp]
  \centering
   \includegraphics[width=0.99\textwidth, trim = {3.3cm 12cm 4.4cm 10cm},clip] {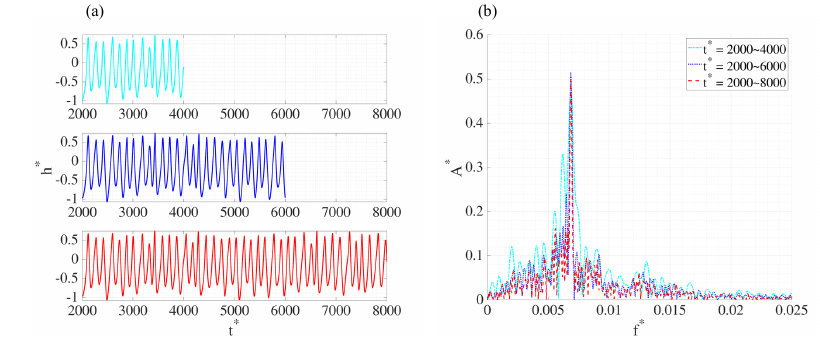}
  \caption{(a) Temporal evolutions  and (b) frequency spectra of interfacial height measured at $x^* = 3$ for different simulation  times. The case is for $\text{Re}=86$ and $\eta=65$ in series B.}   
\label{fig:Time_study_unconfined}
\end{figure}

Additionally, the simulation time is also varied to verify if it is long enough to yield statistically convergence results. The time evolutions and frequency spectra for the interfacial height at two different locations for $\text{Re} = 86$ and $\eta=8, 65$ are shown in Figs.~\ref{fig:Time_study_confined} and \ref{fig:Time_study_unconfined}. The spectra computed based on three different time durations were shown and it can be seen that a long duration will improve the spectrum resolution and to identify the dominant frequency. It should be noted that the transition time for the two-phase mixing layer to fully develop is excluded in the calculation for the frequency spectra. The results here show that the simulation time about $t^*=6000 $ is sufficient.

\end{appendix}


\end{document}